\documentclass{aa}    

\usepackage{graphicx}
\usepackage{epstopdf}
\usepackage{txfonts}
\usepackage{subeqnarray}
\usepackage{natbib} 
\usepackage{version}
\usepackage{rotating}

\begin{document}

\newcommand{\hi}{{\sc Hi}}

\title{An accurate measurement of the anisotropies and mean level of the cosmic infrared background at 100~$\mu$m and 160~$\mu$m}
\author{Aur\'elie P\'enin \inst{1}
\and Guilaine Lagache\inst{1}
\and Alberto Noriega-Crespo\inst{2}
\and Julien Grain\inst{1}
\and Marc-Antoine Miville-Desch\^enes\inst{1}
\and Nicolas Ponthieu\inst{1}
\and Peter Martin \inst{3}
\and Kevin Blagrave \inst{3}
\and Felix J. Lockman \inst{4}}
\institute{Institut d'Astrophysique Spatiale, b\^atiment 121, Universit\'e Paris-XI, 91405 Orsay, France 
\and Spitzer Science Center, California Institute of Technology, MS 100-22, Pasadena, CA 91125
\and Canadian Institute for Theoretical Astrophysics, 60 St George Street, Toronto, Ontario, M5S 3H8, Canada
\and National Radio Astronomy Observatory, P.O. Box 2, Green Bank, WV 24944, USA}

\abstract{The measurement of the anisotropies in the cosmic infrared background (CIB) is a powerful means of studying the evolution of galaxies and large-scale structures. These anisotropies have been measured by a number of experiments, from the far-infrared to the millimeter. One of the main impediments to an accurate measurement  on large scales ($\lesssim$1 degree) is the contamination of the foreground signal by Galactic dust emission.}
{Our goal is to show that we can remove the Galactic cirrus contamination using \hi~data, and thus accurately measure the clustering of starburst galaxies in the CIB.}
{We use observations of the so-called extragalactic ELAIS N1 field at far-infrared (100 and 160~$\mu$m) and radio (21 cm) wavelengths. We compute the correlation between dust emission,  traced by far-infrared observations, and \hi~gas traced by 21 cm observations, and derive dust emissivities that enable us to subtract the cirrus emission from the far-infrared maps. We then derive the power spectrum of the cosmic infrared background anisotropies, as well as its mean level at 100 $\mu$m and 160 $\mu$m.}
{We compute dust emissivities for each of the \hi-velocity components (local, intermediate, and high velocity). Using IRIS/IRAS data at 100~$\mu$m, we demonstrate that we can use the measured emissivities to determine and remove the cirrus contribution to the power spectrum of the cosmic infrared background on large angular scales where the cirrus contribution dominates.We then apply this method to Spitzer/MIPS data for 160~$\mu$m. We measure correlated anisotropies at 160~$\mu$m, and for the first time at 100~$\mu$m. We also combine the \hi~data and Spitzer total power mode absolute measurements to determine the cosmic infrared background mean level at 160 $\mu$m. We find $B_{160}=0.77\pm$0.04$\pm$0.12 MJy/sr, where the first error is statistical and the second one systematic. Combining this measurement with the $B_{100}/B_{160}$ color of the correlated anisotropies, we also derive the cosmic infrared background mean at 100~$\mu$m, $B_{100}=0.24\pm$0.08$\pm$0.04 MJy/sr. This measurement is in line with values obtained with recent models of infrared galaxy evolution and Herschel/PACS data, but is much smaller than the previous DIRBE measurements.}
{The use of high-angular resolution \hi~data is mandatory to accurately differentiate the cirrus from the cosmic infrared background emission. The 100~$\mu$m IRAS map (and thus the map developed by Schlegel and collaborators) in such extragalactic fields is highly contaminated by the cosmic infrared background anisotropies and hence cannot be used as a Galactic cirrus tracer.}
 
\titlerunning{An accurate measurement of CIB mean and anisotropies}
\maketitle
\section{Introduction}

Starburst (SB) galaxies are known to have had an important role in galaxy formation and evolution throughout the whole history of the Universe. In the far-infrared (FIR) and sub-millimeter, observations are limited by extragalactic confusion: details on small spatial scales are lost in the noise because of the poor angular resolution of the instruments. As a result, unresolved starburst galaxies form the cosmic infrared background (CIB) \citep{1996A&A...308L...5P,1998ApJ...508..123F, 1999A&A...344..322L}, which peaks at around 200 $\mu$m. In the mid-infrared, a large fraction of the CIB has been resolved into individual sources: \citet{2004ApJS..154...70P}, for instance, resolved 70 \% of the 24 $\mu$m background. In the FIR, before the advent of the Herschel telescope, a smaller fraction has been resolved: with Spitzer, \citet{2004ApJS..154...87D} resolved 23\% and 7\% of the CIB at 70 and 160~$\mu$m, respectively. \citet{2006ApJ...647L...9F} managed to resolve 60\% of the CIB at 70 $\mu$m using a very deep but small field, which was thus limited by cosmic variance. More recently, \citet{2010AA...518L..30B} integrated counts coming from Herschel/PACS data at 100 and 160 $\mu$m and resolved $\sim45\%$ and $\sim52\%$ of the CIB, respectively. At longer wavelengths, \citet{2010AA...518L..21O} directly resolved 15\%, 10\%, and 6\% of the CIB at 250, 350, and 500 $\mu$m,respectively, using Herschel/SPIRE data. Confusion can be circumvented by the use of statistical methods. For instance,  by stacking 24 $\mu$m sources, \citet{2006A&A...451..417D} were able to resolve a large fraction of the CIB at 70 $\mu$m and 160 $\mu$m and \citet{2010AA...518L..30B} increased their fractions from 45$\%$ to 50$\%$, and from 52$\%$ to 75$\%$ at 100~$\mu$m and 160~$\mu$m, respectively.
Using a P(D) approach, \citet{2011AA...532A..49B} were able to obtain still larger fractions of 65\% and 89\%, at 100 and 160~$\mu$m, respectively;  P(D) derived counts of \citet{2010MNRAS.409..109G} account for 64, 60, and 43\% of the CIB at 250, 350, and 500~$\mu$m, respectively. These results imply that the sources detected at 24~$\mu$m constitute the bulk of the CIB around its peak. \citet{2006A&A...454..143C} showed that galaxies that dominate the emission at 24~$\mu$m become more and more luminous and massive as the redshift increases starting from luminous infrared galaxies (LIRGs) with $10^{11}L_{\odot}<L_{IR}<10^{12}L_{\odot}$ at $0.8<z<1.2$ with intermediate mass, to ultra-luminous infrared galaxies (ULIRGs) with $10^{12}L_{\odot}<L_{IR}<10^{14}L_{\odot}$ that dominate at $z>2$ and have masses $> 10^{11}M_{\odot}$.\\
The clustering of galaxies that make up the CIB can be characterized by its anisotropies. This clustering was first detected at 160~$\mu$m with Spitzer \citep{2007ApJ...665L..89L, 2007A&A...474..731G}, and then measured at 250, 350, and 500~$\mu$m using BLAST data \citep{2009ApJ...707.1766V}. All three of these sets of data enabled the detection of an excess of signal on intermediate spatial scales caused by  the clustering of starburst galaxies which enabled them to derive the linear bias parameter $b$ that relates the density fluctuations of luminous matter to those of dark matter (DM). \citet{2007ApJ...665L..89L} measured a value of  $b=2.4\pm0.2$ while \citet{2009ApJ...707.1766V} obtained $b=3\pm0.3$. The difference may be due to selection effects. At longer wavelengths, higher redshift SB galaxies are probed (\cite{2005ARA&A..43..727L}, \cite{2008A&A...481..885F}) and at these higher redshifts, SB galaxies are a highly biased tracer of the underlying dark matter density field. They indeed formed in very massive DM halos early in the history of the Universe. \citet{2008MNRAS.383.1131M} derived the two-point correlation function of 24~$\mu$m selected sources divided into two redshift bins ($0.6<z<1.2$ and $z>1.6$), finding that these SB galaxies are strongly clustered and embedded in DM halos of $\simeq10^{12.8}M_{\odot}$ for the high $z$ sample and $\simeq10^{11.8}M_{\odot}$ for the low $z$ one. \citet{2010AA...518L..22C} computed the angular correlation function with Herschel/SPIRE data. They found that 250 $\mu$m sources are embedded in DM halos of $\sim10^{12}M_{\odot}$ at $<z>\sim$2.1, whereas bright 500~$\mu$m sources reside in more massive halos, $\sim10^{13}M_{\odot}$, at $<z>\sim$2.6. The CIB anisotropy measurements in the FIR and submillimeter were followed by those of \citet{2010ApJ...718..632H} at 1.3 mm and 2~mm with the South Pole Telescope and by the \citet{2010AAS...21538407F} at 1.4 and 2~mm with the Atacama Cosmology Telescope (\citet{2010AAS...21538408D}). More recently, \citet{2011A&A...536A..18P} derived CIB power spectra from 10\arcmin~to 100\arcmin~simultaneously at 350, 550, 850, and 1380 $\mu$m in six high-Galactic latitude fields. \citet{2011Natur.470..510A} extended the measurements to smaller angular scales, using Herschel/SPIRE observations of the Lockman-hole field at 250, 350, and 500~$\mu$m. These measurements allow us to start to refine the analysis of the clustering properties of galaxies responsible for the CIB, and its cosmic evolution to high redshift (z$\sim$3-4).\\
 The far-infrared and submillimeter emission of the Galactic cirrus interferes with the detection and measurement of the CIB. This emission dominates the power spectrum of the anisotropies on large spatial scales, the exact scale depending on the wavelength and the selected field. To remove the cirrus contribution, data at other wavelengths are usually used. For instance, IRIS maps of  reprocessed IRAS maps \citep{2005ApJS..157..302M} at 100~$\mu$m can be used to determine the power spectrum of the cirrus on large scales. However, as shown in this paper, these data also contain CIB anisotropy and therefore a clustering signal from SB galaxies. To constrain more accurately the contribution of the Galactic cirrus, an external tracer is needed. The most effective one for dust emission in the diffuse sky is neutral hydrogen. In this paper, we show that \hi~data can be used to remove the cirrus contamination from the 100~$\mu$m and 160~$\mu$m maps in order to measure the CIB intensity and the power spectrum of the CIB anisotropy. This method based on template removal to separate the cirrus and CIB components, was also successfully applied to the $\sim$140 square degrees of the very diffuse high-latitude sky observed by both Planck/HFI (350~$\mu$m to 3~mm) and the Green Bank Telescope (21~cm \hi) \citep{2011A&A...536A..18P}. \\
 The paper is organized as follows : we present the data in Sect. 2. In Sect. 3, we briefly describe the astrophysical components and derive the instrumental noise component of the map power spectrum. In Sect. 4, we compute the Galactic cirrus contamination. The cirrus map is then subtracted from the IR maps to obtain the CIB-dominated maps from which we then estimate the power spectrum at both 100~$\mu$m and 160~$\mu$m (in Sect. 5). In Sect. 6, we estimate the mean of the CIB at 160~$\mu$m using the total power mode of Spitzer, and at 100~$\mu$m using the $B_{100}/B_{160}$ color of the measured correlated CIB anisotropies. We then present our conclusions in Sect. 7.
 
\section{Data}\label{par:data_hi_IRIS}
We focus our analysis on the ELAIS N1 field ($\ell,b$)= (85.33$^o$,44.28$^o$), which is part of the Spitzer Wide-Area Infrared Extragalactic Legacy Survey (SWIRE). It covers about ten square degrees and was observed by the Multiband Imaging Photometer for Spitzer (MIPS) at 160~$\mu$m,  by the GBT at 21 cm and by IRAS at 100 and 60~$\mu$m.

\begin{figure}\centering
\includegraphics[scale=0.45]{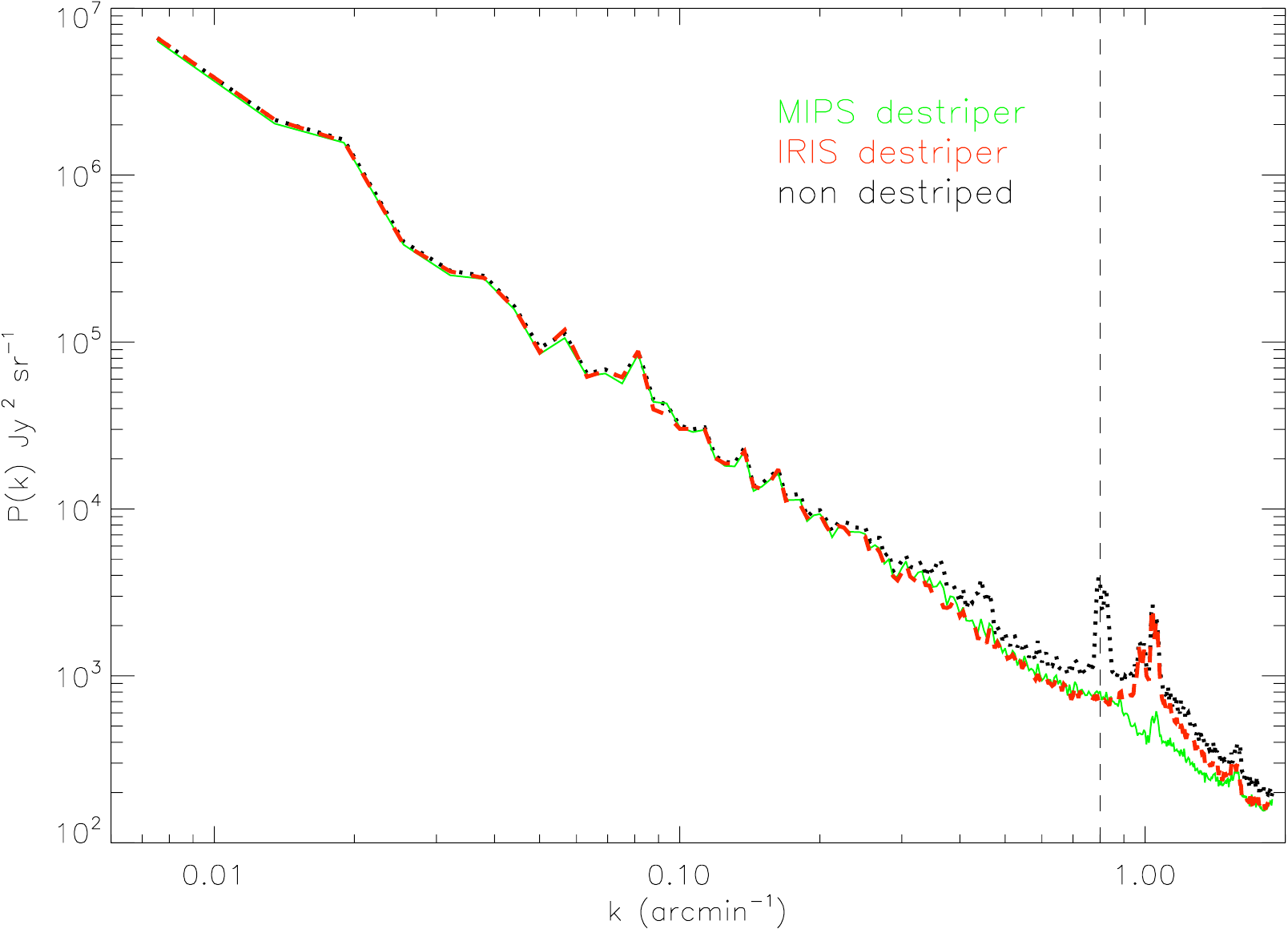}
\caption{The dotted black line shows the power spectrum of the non destriped map, the dashed red curve is the power spectrum of the map destriped with the "IRIS destriper", and the green line shows the power spectrum of the map destriped by "the MIPS destriper". The vertical dashed line shows our angular scale cut of $<$~0.8 arcmin$^{-1}$.}
 \label{fig:destriped}
\end{figure}

\begin{figure}[]\centering
\includegraphics[scale=0.40]{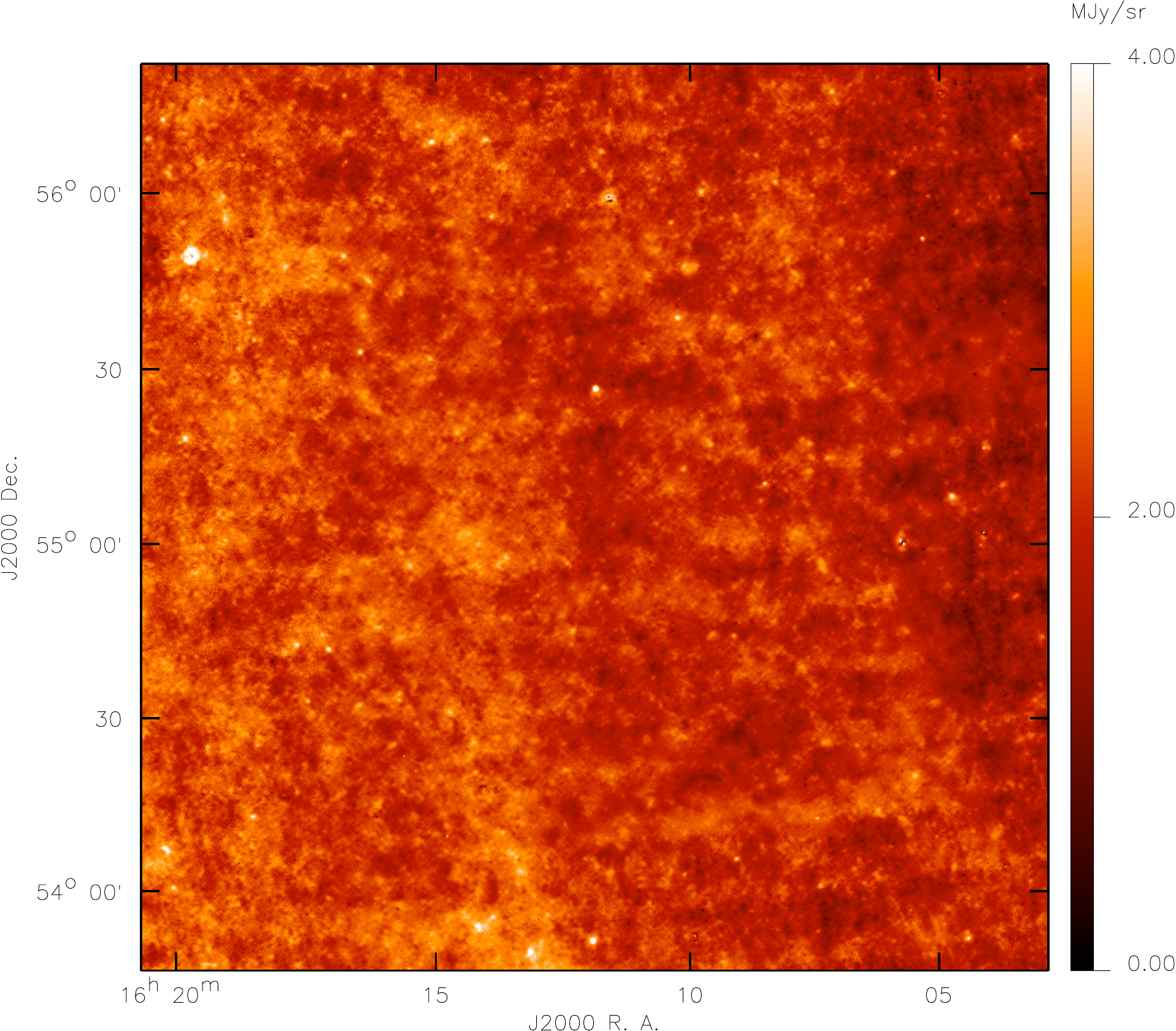}
\caption{The final source-subtracted 160~$\mu$m map centered on ELAIS N1 used to compute the power spectrum. Units are MJy/sr.}
 \label{fig:mips_map}
\end{figure}

 \begin{figure}
\begin{minipage}[]{.35\linewidth}
  \centering\includegraphics[scale=0.21]{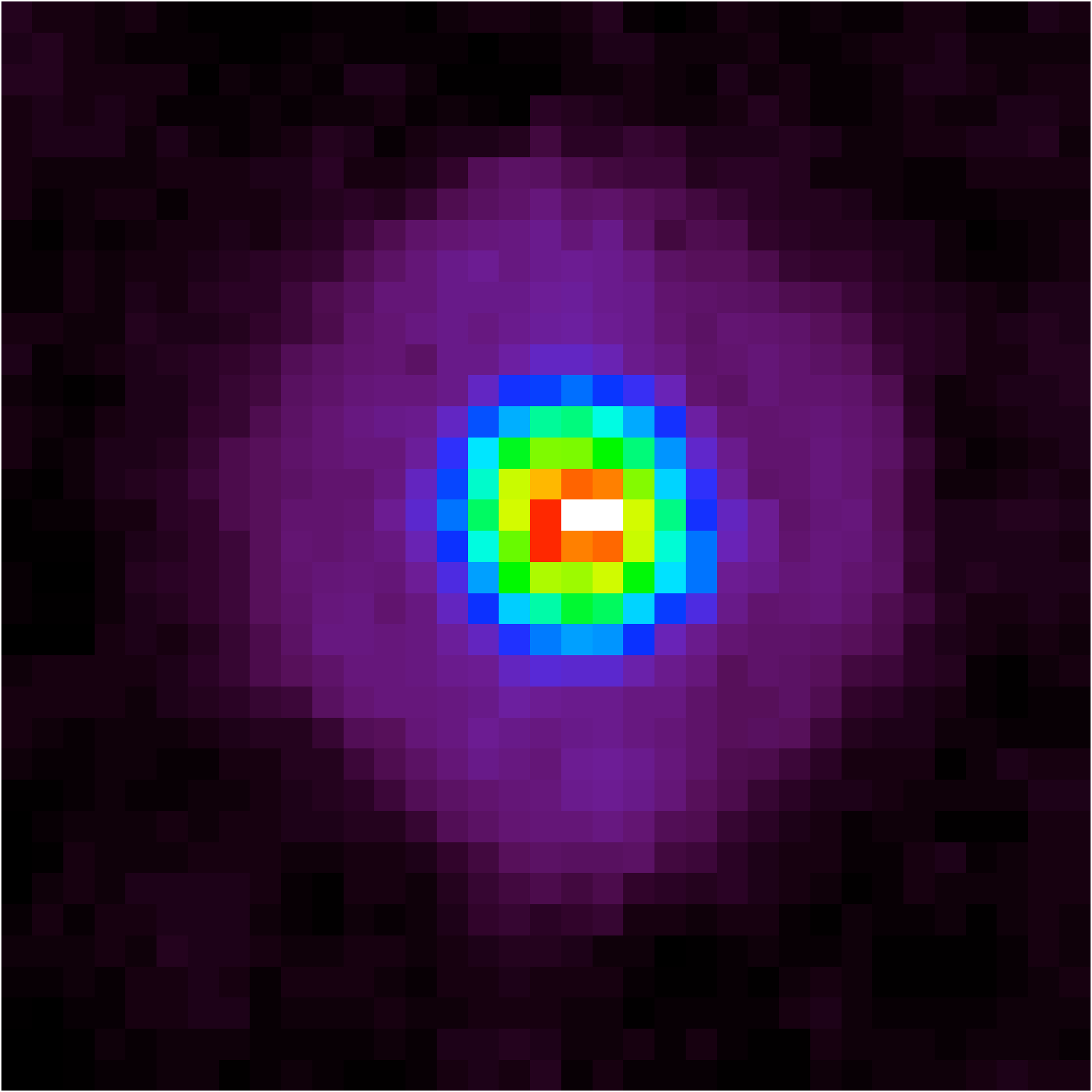}
\end{minipage}
\hfill
\begin{minipage}[]{.65\linewidth}
  \centering\includegraphics[scale=0.30]{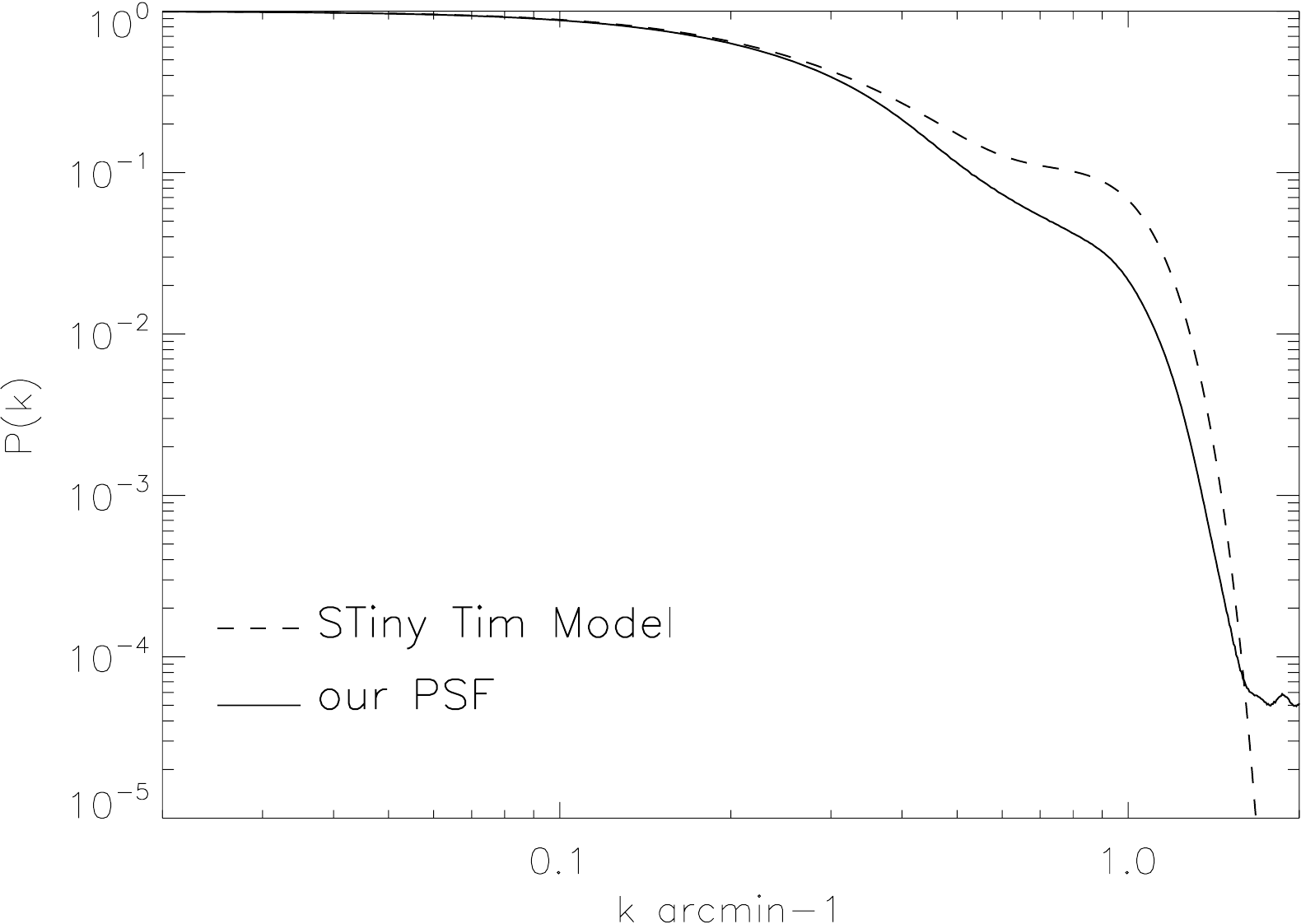}
\end{minipage}
\caption{Left: point spread function (PSF) of MIPS 160~$\mu$m computed using an oversampled map. Right:  PSFs power spectra. The continuous line is our PSF and the dashed line is the power spectrum from the STiny Tim Model (\cite{2005AJ....129.1008K}, STinyTim, v1.3; Krist, 2002).}
 \label{fig:psf}
\end{figure}

\subsection{MIPS 160 $\mu$m}\label{par:mips}
The MIPS observations at 160 $\mu$m~were taken as part of the SWIRE legacy program \citep{2003ApJ...592..804L} in 2004 during two epochs separated by
six months (late January and late July). The data were taken using the scan
mode and the medium scan rate, in about three-degree long strips and an offset
between the return and forward scans of 148\arcsec. This observing strategy
produced a map with a median depth coverage of 8 (times 4 seconds per
frame = 32 sec integration time). However, owing to the dead readout of the
160~$\mu$m~array \citep{2007PASP..119.1038S} the coverage was inhomogeneous,
being as low as two or as high as ten in some regions of the map,
usually along the scanning direction and overlapping independent regions.\\
During the observations of the first epoch, the Spitzer Observatory went
into stand-by mode (Jan 25th). This meant that some of the 160 $\mu$m~observations
were affected by a slightly warmer telescope during the recovery
phase, with a temperature of around 6 K rather than the standard $\sim 5.6$ K. We mitigated this effect in the data by applying a small offset determined from the nearby unaffected regions (an overlap correction). \\
The fact that the 160 $\mu$m data were taken during two different epochs makes the
processing and creation of the final mosaic relatively straightforward.
We used the standard basic calibrated data (BCDs) from the Spitzer Science
Center, and removed their prediction of the zodiacal light as a function of time and space 
from each BCD and carried out an overlap correction \citep{2008PASP..120.1028M}.
The maps were created using the native pixel scale of 16 arcsec/pixel, which
preserves the diffuse emission. These final mosaics show no evidence of any defects other than low intensity stripes left as an artifact of the scanning observation. These stripes introduce two peaks at $k\sim$1~arcmin$^{-1}$ into the power spectrum of the map, as shown in Fig.~\ref{fig:destriped}. The effect of these stripes was mitigated by applying a destriping filter using ridgelets (Ingalls et al. 2011). Our tests of the photometry of the maps and their sources showed that the destriping method preserves the flux to within 5\% of the original values. However, the power spectrum of the destriped map shows a little loss of power on small scales (k$>$0.2~arcmin$^{-1}$), as can be seen by comparing the green line with to the black dotted line in Fig.~\ref{fig:destriped}. To investigate whether any astrophysical information had been removed or only power contained in the  stripes, we destriped the map using the destriper algorithm developed for IRIS, which has been shown to prevent the removal of astrophysical signal \citep{2005ApJS..157..302M}. 

\begin{figure}[]\centering
\includegraphics[scale=0.4]{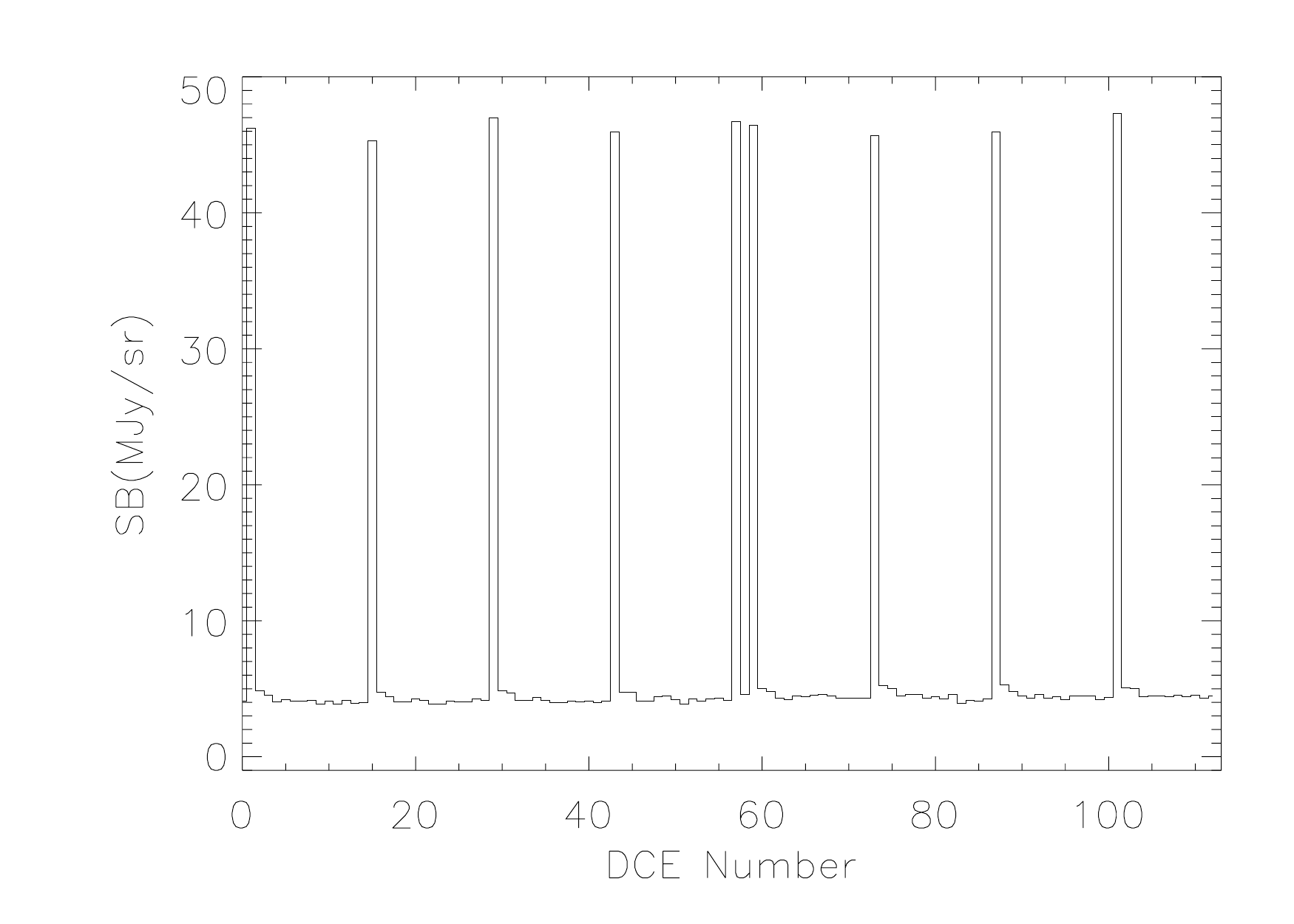}\caption{The complete data sequence of one photometric observations of the ELAIS N1 field at 160 $\mu$m, with the median values of each frame in surface brightness units (MJy/sr) as a function Data Collection Event (DCE) number (essentially time).  The highest values correspond to the calibration stim flashes.}
\label{fig:101}
\end{figure}

\begin{figure}[]\centering
 \includegraphics[scale=0.4]{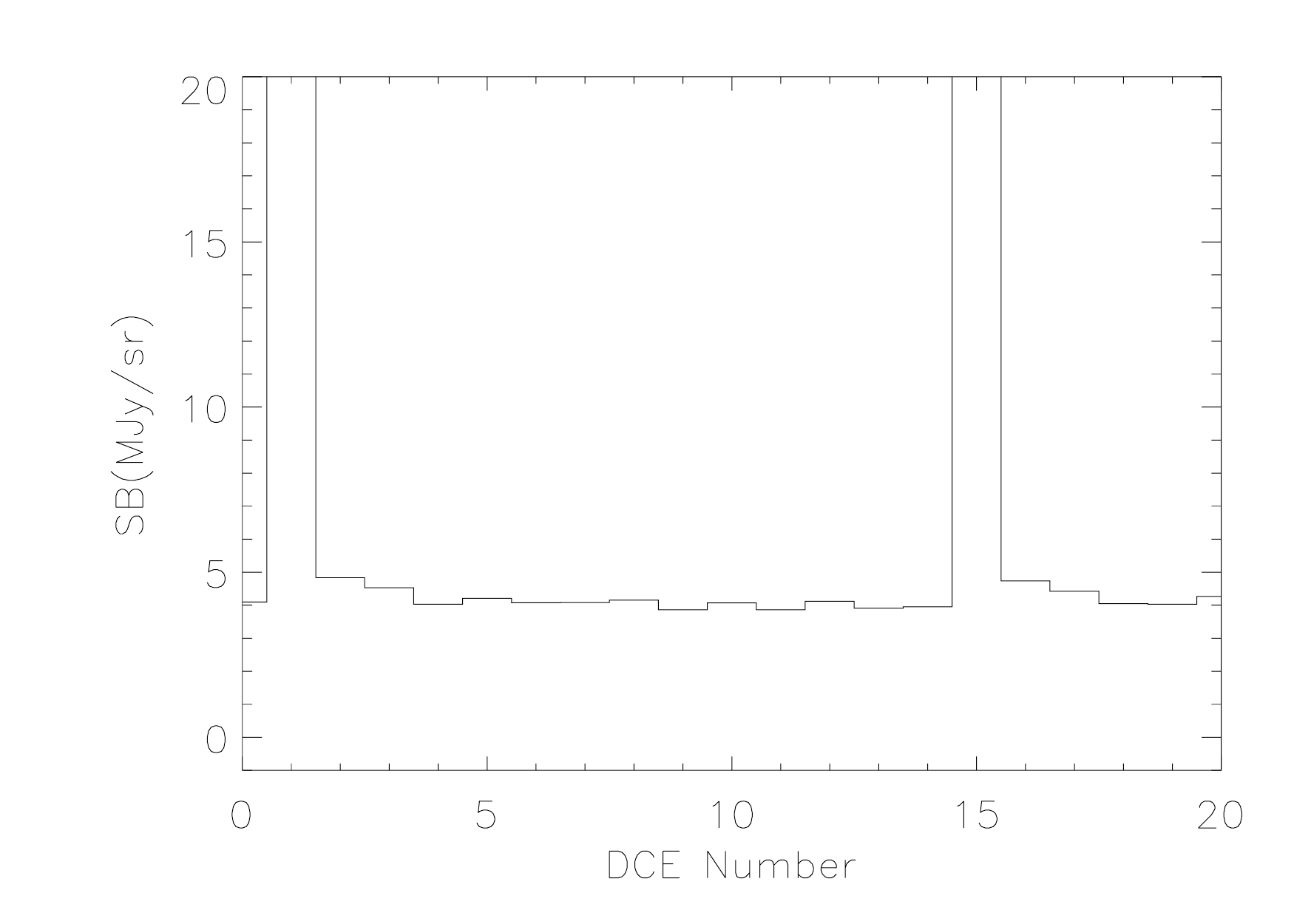}
\caption{A zoomed image of the first 20 DCEs of the Photometic observation, showing the small exponential decay after a stim flash lasting about 2-3 DCEs.}
\label{fig:102}
\end{figure}

The power spectrum of that map is shown by the red curve in Fig. \ref{fig:destriped}. It is in good agreement with the green curve up to $k\simeq$0.8~arcmin$^{-1}$, suggesting that the ``ridgelet destriping'' did not remove any astrophysical signal. At higher $k$, we found that the IRIS destriper fails to remove residual stripes as it was not developed for such high-angular resolution data. We therefore use the map destriped with ridgelets. To limit point spread function (PSF) uncertainties (see below) we consider only scales $<$0.8~arcmin$^{-1}$, hence disregard the residual stripe at $k\sim1$~arcmin$^{-1}$.\\
The identification of sources and their extraction from the Spitzer
long-wavelength images (70 and 160$\mu$m) has been carried out on
filtered mosaics using the standard methods applied to the `deep fields' (see e.g. \cite{2006AJ....131..250F}, \cite{2006ApJ...647L...9F}, \cite{2009AJ....138.1261F}). Filtering has been performed to remove the extended emission, thus enhance the detectability of faint point sources.
Since we wish to preserve the
background/foreground emission, we carried out the source removal on our original mosaics.
The PSF is well-sampled in MIPS data: the full width halh maximum (FWHM) of the 160 $\mu$m PSF is 40\arcsec compared to the pixel size of 16\arcsec. We thus selected {\tt Starfinder}, which uses a PSF fitting algorithm, to perform our source extraction (Diolatti et al. 2000). {\tt Starfinder} also has the advantage that it evaluates the background over the entire image,
in the course of  the iterative fitting of the individual sources: this leads to smaller
residuals \citep{2004ApJS..154...66M}, even compared to those of the standard
Spitzer source extraction software {\tt MOPEX/APEX} \citep{2005PASP..117.1113M}. We detected sources down to $S_{160}=25\pm5$~mJy and removed them from the map. Even though the source catalog is highly incomplete at low fluxes, removing sources down to very faint fluxes allows us to lower the Poisson part of the power spectrum and accurately measure the CIB correlated part. Fig. \ref{fig:mips_map} shows the source-subtracted map that we use in our study.\\

\begin{figure}[]\centering
 \includegraphics[scale=0.4]{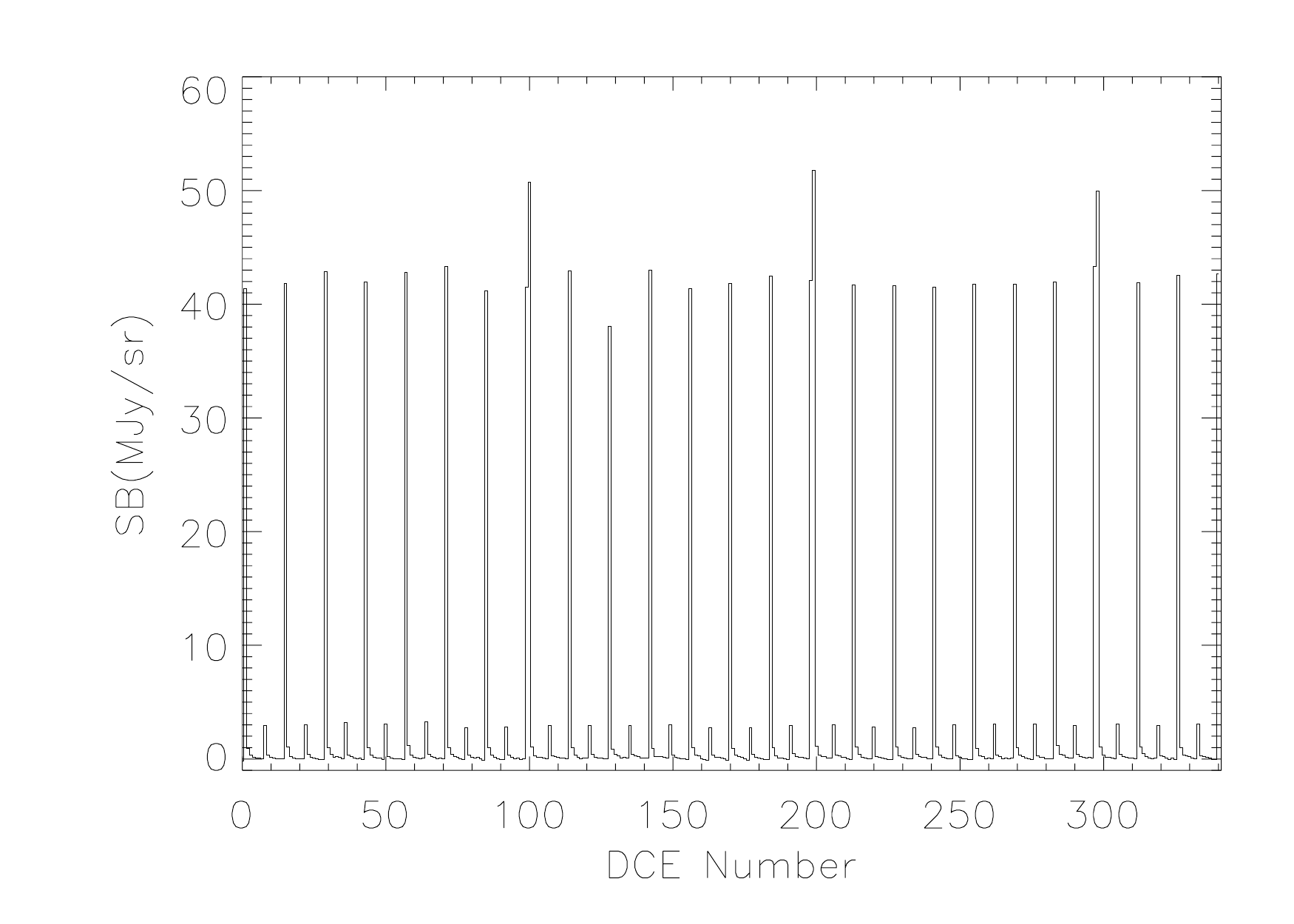}
\caption{As Fig. \ref{fig:101}, but for the total power mode observation. Once again the largest values correspond to the stim flashes, with the sky measurement being shown in-between them.}
\label{fig:103}
\end{figure}

\begin{figure}[]\centering
 \includegraphics[scale=0.4]{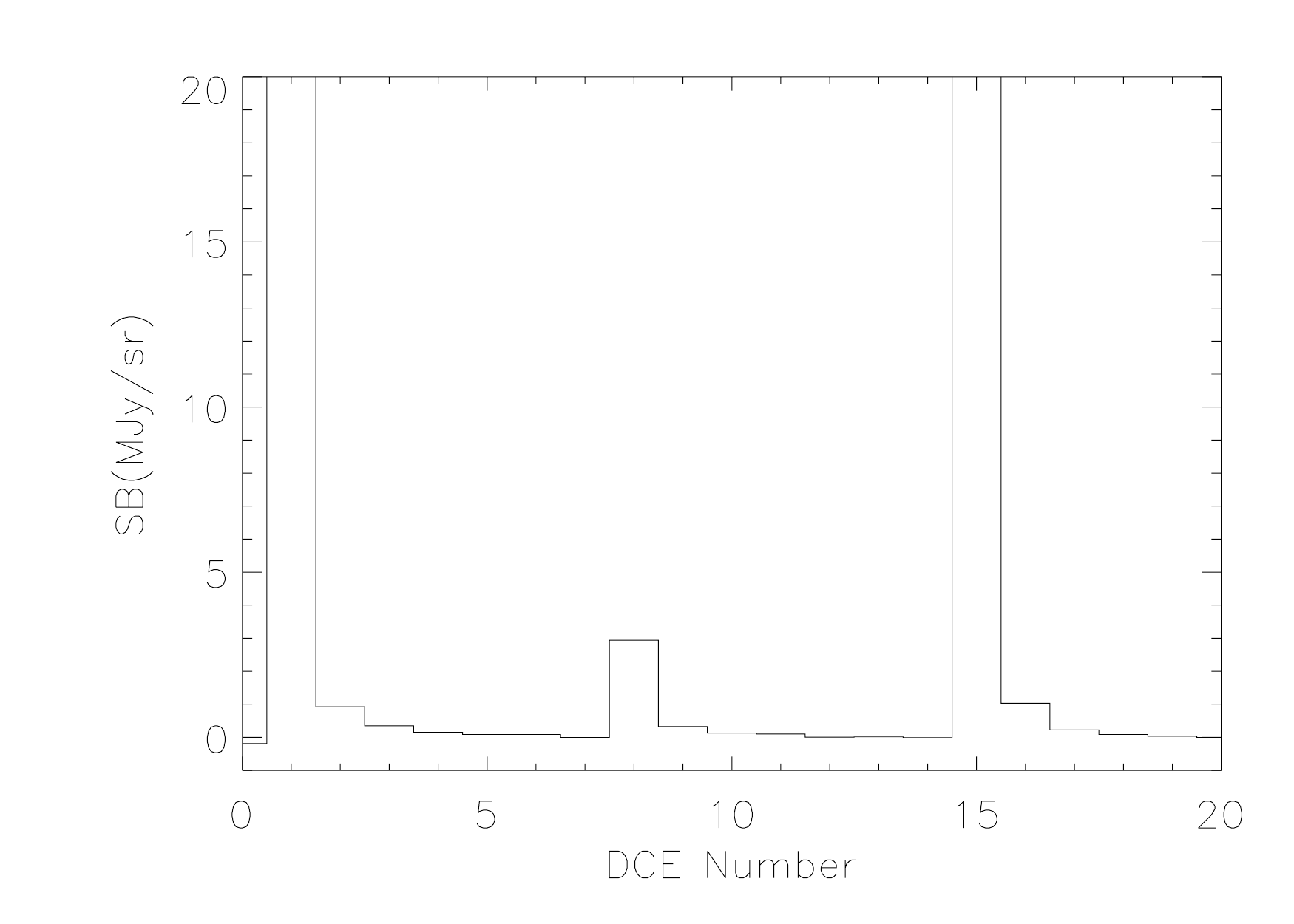}
\caption{A zoomed image of the first 20 DCEs of the total power mode observation, showing as well the exponential decay, the sky measurement (DCE=8), small decay, and the next calibration stim.}
\label{fig:104}
\end{figure}

The effective PSF of the observed data is quite sensitive to the observed field. We therefore had to determine it directly in the map rather than use a prior estimate. We computed the PSF by stacking the brightest sources of an oversampled map (7.2 arcsec/pix). The oversampling was necessary to obtain an accurate profile of the PSF. We extracted sources at 5$\sigma$ and retained only those with $S_{160}>600$ mJy. The left panel of Fig. \ref{fig:psf} displays the extracted PSF and the right panel its power spectrum compared to that of the STiny Tim Model\footnote{http://ssc.spitzer.caltech.edu/dataanalysistools/\\tools/contributed/general/stinytim/}(\cite{2005AJ....129.1008K}, STinyTim, v1.3; Krist, 2002). The discrepancy between the power spectra shows the need to determine the PSF in our data. We estimated the errors in the PSF, using different oversamplings, bright source flux cuts, and other extragalactic fields observed with the same scanning strategy. We found that our measurement is highly reproducible up to $k\simeq$0.8~arcmin$^{-1}$. At higher $k$, we have differences that can be as large as  35\% which prevented us  from accurately measuring the  CIB power spectrum. We will thus analyze the power spectra only for $k\le$0.8~arcmin$^{-1}$.

\begin{figure*}[]
\includegraphics[scale=0.22]{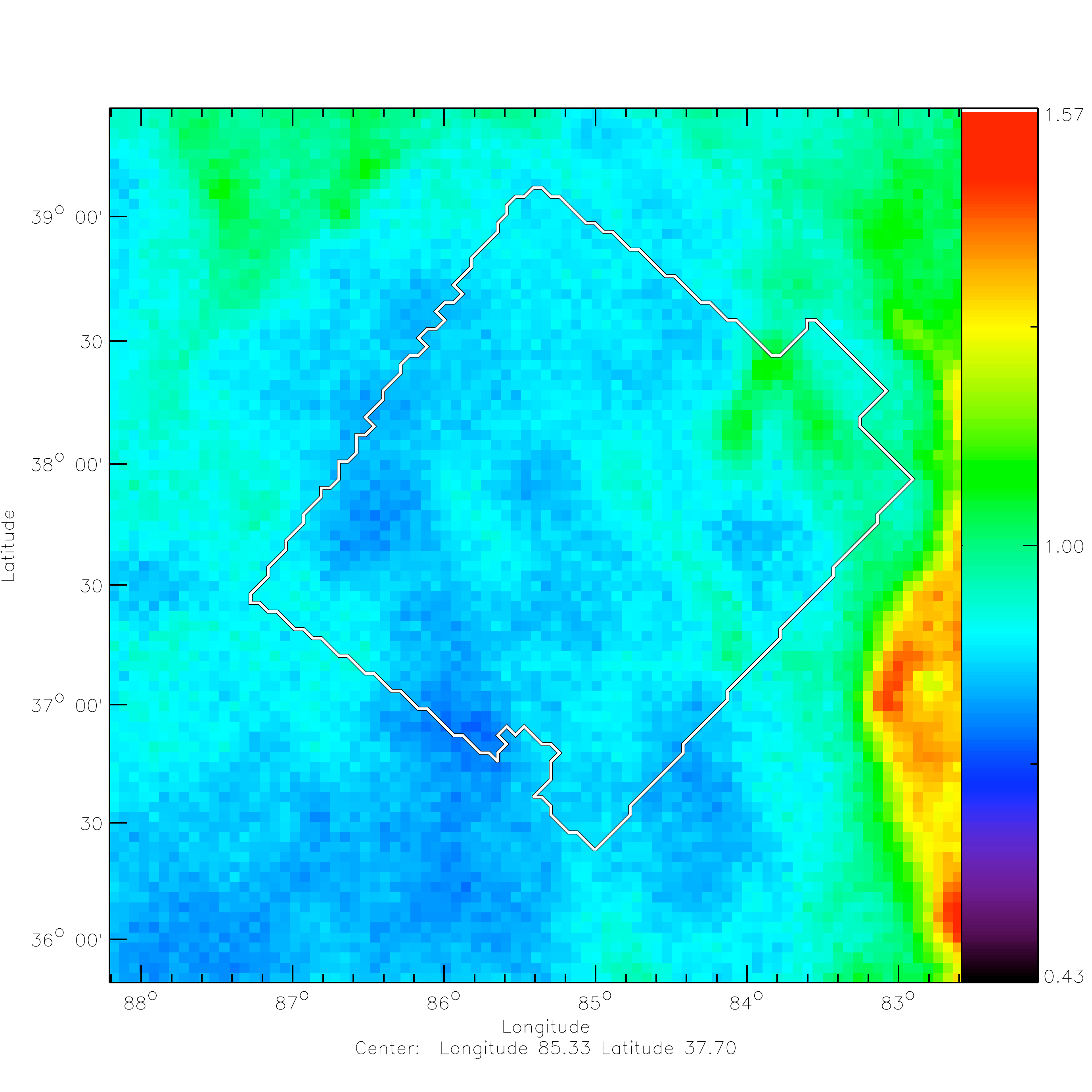}\includegraphics[scale=0.22]{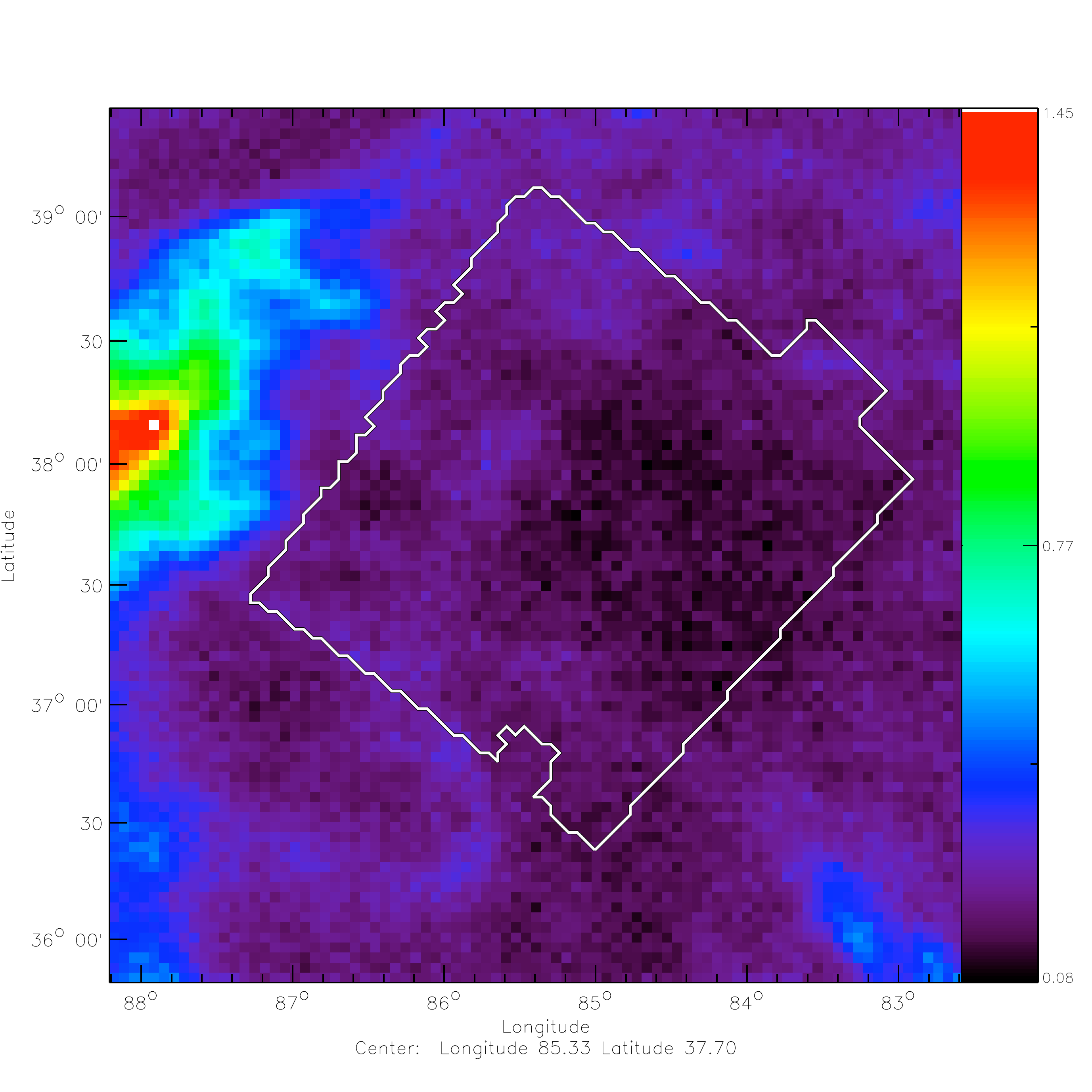}\includegraphics[scale=0.22]{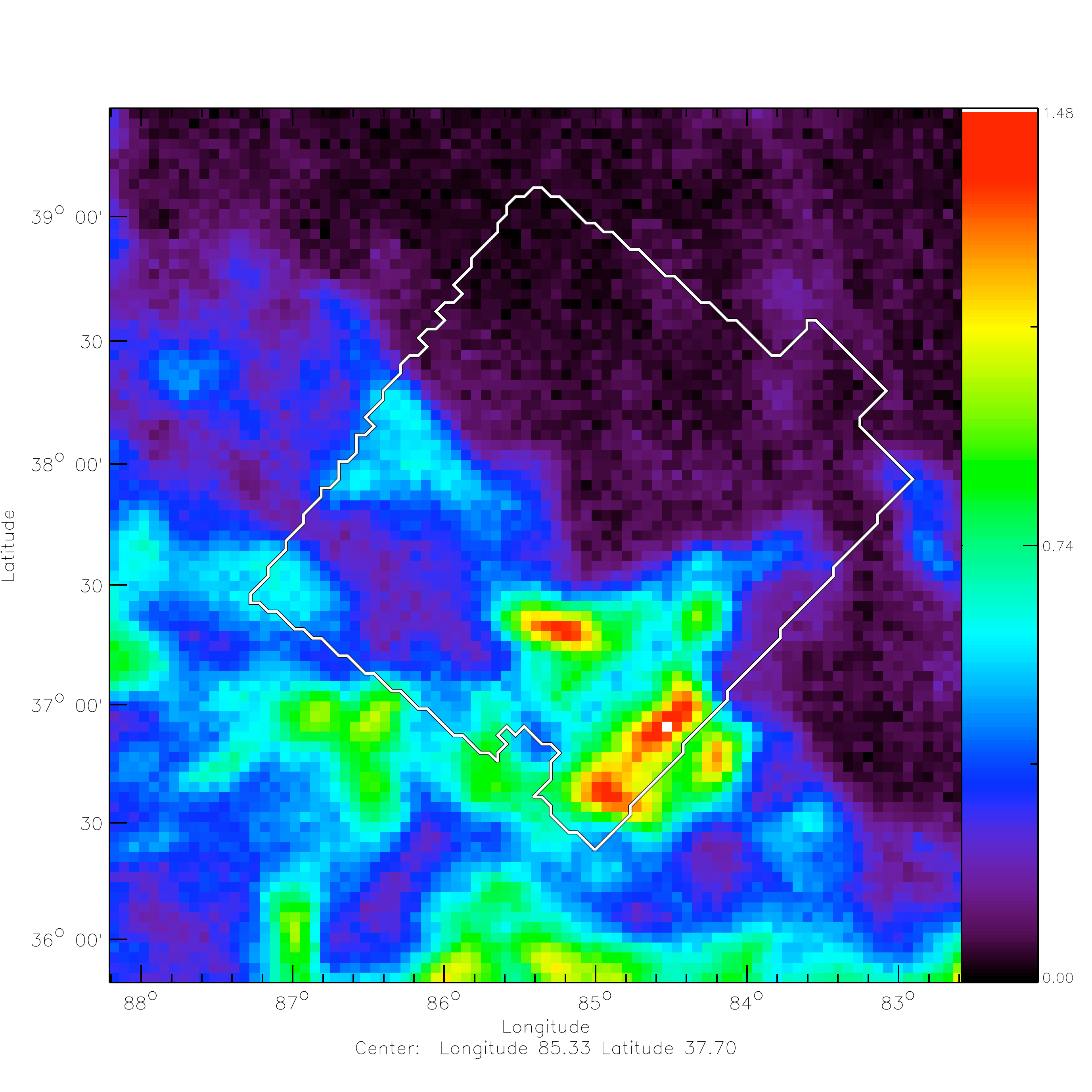}
\caption{ELAIS N1/GBT column densities $N_{HI}$, in unit of 10$^{20}$ atoms/cm$^2$. From {\it left} to {\it right}: the column density of the local component with -14~km/s $<V_{LSR}<$ 43~km/s, the IVC with -79~km/s $<V_{LSR}< $-14~km/s and the HVC with -163~km/s $<V_{LSR}< $-79~km/s. The white contour in each image shows the field covered by \textit{Spitzer} (ELAIS N1/MIPS field).}
 \label{fig:GBT}
\end{figure*}

\subsection{MIPS 160 $\mu$m total power modes}\label{par:tpmobs}
In our analysis, we used of total power mode (TPM) MIPS observations.
A TPM observation at 70~$\mu$m and 160~$\mu$m~is merely another example of a 
photometry mode observation in terms of its processing from raw data to calibrated images 
(basic calibrated data or bcd). As in the case of photometry, the calibration was 
directly tied to the stim flashes that are measured approximately every two
minutes. In the standard photometric observations, the scan mirror  
places the source/sky on different positions of the array(s), taking an image
of a frame at each corresponding position. Each frame is calibrated using
the stim flashes, and goes through several steps as described in 
\citet{2005PASP..117..503G}, described by equations (4) to (16), with two steps being 
fundamental and applied to all the source frames, a dark subtraction and 
an illumination correction, i.e. division by a sophisticated flat.
The total power mode differs in one fundamental way from the photometry mode, in that instead
of using the scan mirror to move only the source around the array, it performs
an intermediate step of placing the mirror on an internal dark, an `absolute
reference' frame, between the source observations. This internal reference dark
is inside the cold instrument, i.e. a temperature of less than 1.5 K
(recall that the Spitzer telescope cryogenic minimum temperature was 5.6 K). 
Laboratory and flight tests have shown that the attenuation is at least a factor ten 
when the scan mirror was placed at the 160~$\mu$m~dark position. This 'reference' frame is
observed in the same way as the sky or source frames i.e. is calibrated using stim flashes
and undergoes a dark subtraction and illumination correction. As pointed
out by \citet{2005PASP..117..503G} in their Sect.~5.2.1, one problem when flashing a stim
on the 70~$\mu$m and 160~$\mu$m~Germanium arrays is that of a memory effect (`stim flash
latent') soon after the flash. This latency decays exponentially in a way that 
depends on the background light, so it can last 5-20 s and has a peak amplitude
of less than 3$\%$ for 70 $\mu$m and 7$\%$ for 160~$\mu$m. Experiments have shown that
for signals of a few MJy/sr, similar to those in this study, the latencies are
almost completely reduced.
 We note that TPM observations were meant to be performed on diffuse signals and relatively dark
regions of the sky, hence the design of the observing mode was done by 
taking into account this memory or latency effect. A standard 160~$\mu$m TPM observation
places the array between the sky and the 'internal reference' (aka 'internal dark'),
but after each observation the memory effect is allowed to decay by taking a series of
seven frames, and using the frames unaffected by the latency of the final measurement.
Figures \ref{fig:101} through \ref{fig:104} show the time history, as a function of the frame or Data Collection Event (DCE) number,
of the median value for each BCD (i.e. the median over the 2$\times$ 16 160~$\mu$m~array)
from one of photometry and TPM observations carried out at 160~$\mu$m~for the ELAIS N1 field. These
observations were done in a sequence, i.e one after the other.
Fig. \ref{fig:101} and \ref{fig:102}, show the photometric observation; the highest values $\sim$50 MJy/sr 
corresponds to the stim flashes and the rest the sky measurement. The zoomed image shows 
the first 13 DCEs in-between the two stims (DCE=1 and DCE=15) approximately two minutes apart. 
We note how pairs of frames after the stims are affected by the latency memory effect, 
and that the overall level is $\sim 4.5$ MJy/sr. 
Fig \ref{fig:103} and \ref{fig:104} show the TPM observation, with once again the median values of the individual
BCDs as a function of time (DCE number). The strongest signal is that of the stims and the 
weakest signal that of the source/sky itself. The zoomed image shows the first 20 DCEs,
where DEC=0 corresponds to the sky measurement, DCE=1 the stim, DCE= 2 to 7 
to observations in the "internal dark reference" and DCE=8 is the sky measurement followed by 
another six DCEs and the next calibration stim (DCE=15).  We note that the level of the
DCE=7 is essentially zero, because in the data reduction pipeline this is the 
'reference' measurement that is subtracted from each BCD. We also note that the sky itself
(DCE=8) is at $\sim3$ MJy/sr, which is the final TPM measurement. The comparison
of the TPM with the photometric observation shows that the light contribution of the telescope
at 160 $\mu$m~is $\sim 1.3$ MJy/sr. The level of spurious emission due to the telescope 
background emission at 160~$\mu$m~ is 1.0$\pm$0.2 MJy/sr.\\
The absolute calibration of 160~$\mu$m~TPM relies on the standard 160 $\mu$m calibration,
which is based on asteroids, that is tied itself to the 24 and 70~$\mu$m~MIPS absolute
calibration to be internally consistent \citep{2007PASP..119.1038S}.
Finally, we stress that a single DCE at 160~$\mu$m~does not cover the
160 $\mu$m~beam (40\arcsec), hence the TPM mode was designed to move the scan mirror
to cover the beam in one cycle. A standard TPM 160~$\mu$m~observation contains four cycles.
The final product for a single 160~$\mu$m  TPM observation is a 5\arcmin$\times$5\arcmin~
small mosaic, and for our measurements we calculated the mean over such an image.
The two ELAIS N1 TPM observations at 160 $\mu$m took 2386 s with 88 s on source each, 
while the two photometric observations using the enhanced mode took 671 secs with 54 secs
on each source. These numbers illustrate the efficiency of the two modes; for every second on source
at TPM, about seven seconds woth of data are used for calibration and latency decay.

\subsection{\hi~data}\label{par:datahi}
\begin{figure}[]\centering
\includegraphics[scale=0.60]{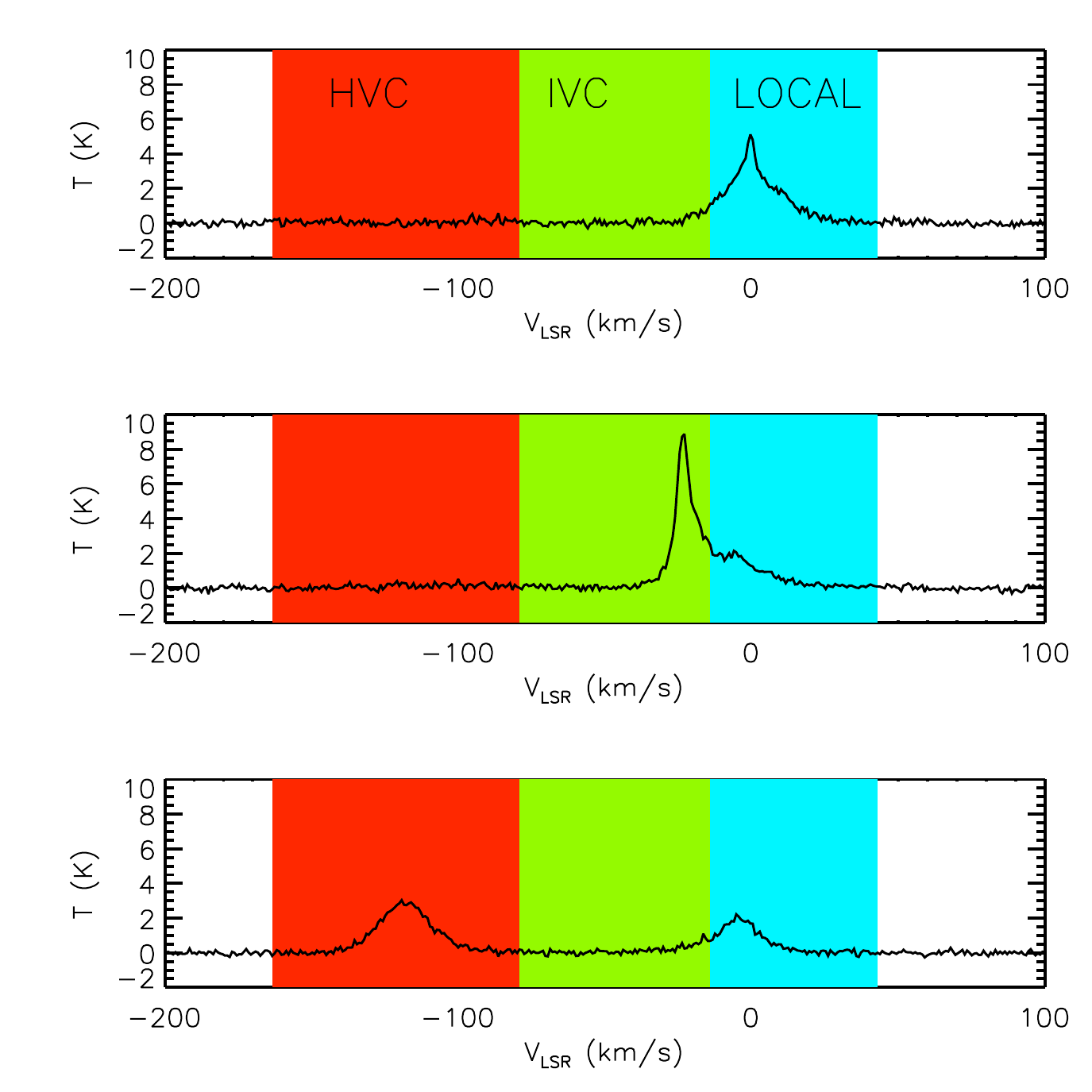}
\caption{\hi~spectra of three different lines of sight, illustrating the three velocity components (local, IVC, and HVC from the top to the bottom panels, respectively).}
 \label{fig:comp_spectra}
\end{figure}

 We used the \hi~25 deg$^2$ data cube (x, y, velocities) centered on ELAIS N1. These data were obtained in 2006 and 2010 with the 100-meter Green Bank Telescope (GBT).  Spectra were measured over a $5^o \times 5^o$ area centered on ($\ell$,$b$) = (85.5$^o$,+44.3$^o$) every 3.5$\arcmin$ in both coordinates. Data were taken by in-band frequency switching yielding spectra with a velocity coverage $-450 \leq V_{LSR} \leq +355$ km s$^{-1}$ and a velocity resolution of 0.80~km.s$^{-1}$.  Spectra were calibrated, corrected for stray radiation, and placed on a brightness temperature ($T_b$) scale as described in \citet{2010ASPC..438..156B} and \citet{2011A&A...536A..81B}.  A third-order polynomial was fit to the emission-free regions of the spectra to remove any residual instrumental baseline.  The final data cube has a root mean square (rms) noise in a single channel of 0.12 K of $T_b$, and an effective angular resolution of $9.4 \arcmin \times 9.1 \arcmin$ in $\ell$ and $b$, respectively.\\
We distinguish three velocity components in the \hi~gas data: the local, intermediate (IVC), and high velocity cloud (HVC). These are shown in Fig. \ref{fig:GBT}. The HVC is centered around -115 km/s and the IVC around -23 km/s, as illustrated in Fig. \ref{fig:comp_spectra}. This figure shows velocity spectra along three lines of sight, each of which is dominated by one component. The IVC and the HVC are clearly seen in the middle and bottom panels.
\subsection{IRIS/IRAS data}\label{par:dataIRIS}
\begin{figure}[]\centering
\includegraphics[scale=0.45]{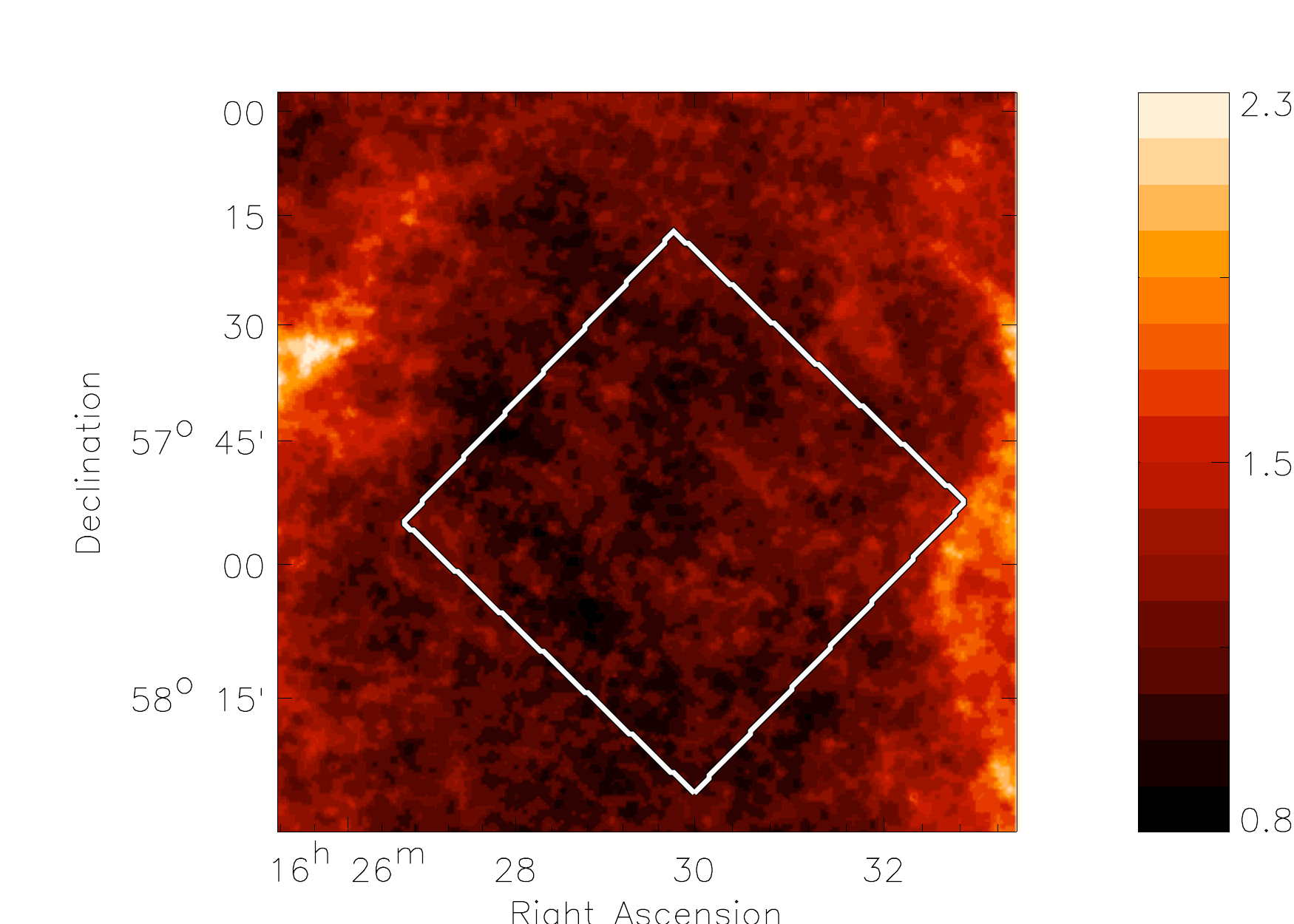}
\caption{Source-subtracted IRIS 100 $\mu$m \textbf{map} centered on ELAIS N1. The area is $\sim25$ square degrees. Units are MJy/sr. The white contour shows the MIPS field.}
 \label{fig:iris_map}
\end{figure}

We used IRIS (re-processed IRAS data) maps at 60 and 100~$\mu$m to measure the emissivities of the dust correlated to the \hi~components and derive the CIB power spectrum at  100~$\mu$m. This new generation of IRAS images was processed using a more reliable zodiacal light subtraction, from a calibration and zero level compatible with DIRBE and a more reliable destriping \citep{2005ApJS..157..302M}. At 100~$\mu$m, the IRIS product also represents a significant improvement on the \citet{1998ApJ...500..525S} maps. IRIS keeps the full ISSA resolution, it includes well-calibrated point sources, and the diffuse emission calibration on scales smaller than one degree was corrected for the variation in the IRAS detector responsivity with scale and brightness.\\
Using IRAS, two full-sky maps (HCON-1 and HCON-2 for hours confirmation) were performed and a last one that covers 75 \% of the sky (HCON-3). The three of them were processed in the same way, including deglitching, checking of the zero-level stability, visual examination for remaining glitches and artifacts, zodiacal light removal, and gain calibration. The three HCONs were then coadded using sky coverage maps to produce the average map (HCON-0). We return to these HCONs later to determine the power spectrum of the instrument noise.\\
The IRIS PSF is assumed to be Gaussian following \citet{2002A&A...393..749M}
\begin{equation}
P(k) = exp\left({-\frac{k^2}{2\sigma_k^2}}\right), 
\end{equation}
where $\sigma_k = 0.065\pm 0.005$ arcmin$^{-1}$ at 100~$\mu$m, which corresponds in real space to a Gaussian function with $\sigma = 1.8\pm 0.1$ arcmin. 
Sources are removed down to a 10$\sigma$ threshold, following the algorithm described in \citet{2005ApJS..157..302M}.
The source-subtracted map at 100~$\mu$m is shown in Fig. \ref{fig:iris_map}.\\
The FWHM of our data are given in Table \ref{tab: FWHM}. We convolved the maps when necessary in order to compare consistent data.\\
\begin{table*}
\begin{center}
\begin{tabular}{*{5}{c}}
\hline \hline 
              & MIPS 160 $\mu$m  & IRIS 100 $\mu$m & IRIS 60 $\mu$m & GBT 21 cm  \\
\hline 
FWHM (arcmin) & 0.64             & 4.3$\pm$0.2     & 4.0 $\pm$0.2   & 9.1\\
\hline 
\end{tabular} 
\caption{Effective measured PSF full width half at maximum (from \citet{2007PASP..119.1038S}, \citet{2005ApJS..157..302M} and \citet{booth} at 160~$\mu$m, 100~$\mu$m and 21~cm, respectively)}
\end{center}
\label{tab: FWHM}
\end{table*}

\section{Power spectrum and error bars}\label{par:pk_and_eb}
\begin{figure}[]
\begin{center}
\includegraphics[scale=0.45]{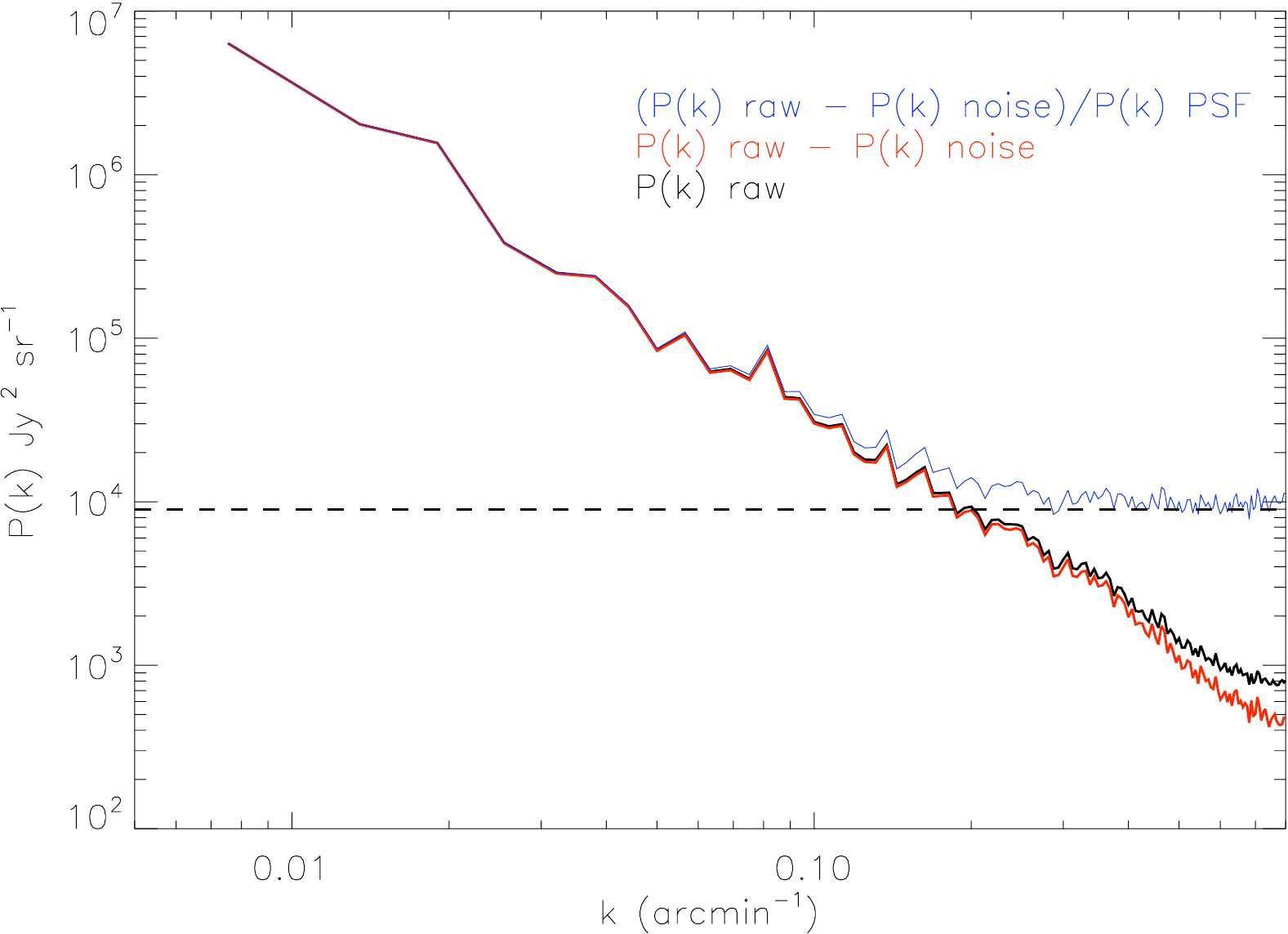}
\caption{The black line shows the raw power spectrum, $P(k)$, of the map at 160 $\mu$m. The red line is the noise-subtracted power spectrum $P(k)-N(k)$. The blue one represents the noise-subtracted power spectrum divided by that of the PSF $(P(K)-N(k))/\gamma(k)$. The dashed line shows the level of Poisson noise, which is $9\times 10^3 $Jy$^2$/sr. For clarity, the error bars are not shown.}
\label{fig:pk_moins_bruit_dec}
\end{center}
\end{figure}

There are several contributions to the power spectrum measured in the FIR. These include the Poisson noise caused by discrete unresolved sources and the clustering of galaxies, which together represent the CIB anisotropies, the Galactic cirrus, and the instrument noise. Assuming that the noise is not correlated to the signal
\begin{equation}
P(k) = \gamma(k)\left[P_{sources}(k)+P_{clus}(k)+P_{cirrus}(k)\right]+N(k),
\end{equation}
where $k$ is the two-dimensional wavenumber and $P_{sources}(k)$, $P_{clus}(k)$, and $P_{cirrus}(k)$ are, respectively, the power spectrum of unresolved sources, the clustering, and the Galactic dust emission. The instrumental noise is represented by $N(k)$, and $\gamma(k)$ is the power spectrum of the PSF of the instrument.\\
The noise power spectrum, $N(k)$ was computed using two independent
maps of the ELAIS N1 field. At 100~$\mu$m, we used the different HCONs (see
Sect. \ref{par:dataIRIS}). At 160~$\mu$m, we used the even and odd BCDs to build two independent maps. The power
spectrum of the difference in the two maps gives an estimate of
$N(k)$. To take account of the inhomogeneous coverage, we applied
the method of \citet{2005ApJS..157..302M}. We subtracted the estimated
$N(k)$ from the raw power spectrum $P(k)$. The level of the Poisson
noise at large $k$ is obtained by dividing the power spectrum by that
of the PSF.\\
The result is shown in Fig. \ref{fig:pk_moins_bruit_dec}. We found that $P_{sources}=9013\pm 100$ Jy$^2$/sr. \citet{2007ApJ...665L..89L} found a slightly higher value of 9848$\pm 120$ Jy$^2$/sr with sources removed with a higher flux (200 mJy). In this study, sources are removed to a lower flux cut, which leads to a lower shot noise level.\\
Statistical errors in power spectra are computed using mock signal plus noise maps that we analyzed with the same pipeline as the data. We derivedthe covariance matrix of this set of power spectra. Its diagonal terms give the errors in each $P(k)$. Errors in the subtraction of the cirrus component are not statistical but systematic and due to errors in the emissivities $\delta \alpha_i$ (see Sect \ref{par:eb2}). Errors in the power spectrum of the cirrus-subtracted map due to the spatial removal are on the order of $(\delta \alpha)^2\times P(k)$, which is negligible compared to the statistical error. \\
We have seen in Sect. \ref{par:mips} that there are uncertainties in the determination of the PSF. We took these into account by using several PSFs (ours, STiny Tim, and one computed from another Spitzer field, the CDFS) in the same pipeline. We derive again the covariance matrix of the set of power spectra. We add these errors in quadrature to statistical errors.\\

\section{The Galactic component}
In previous works, the Galactic contribution to the power spectrum was either assumed to be proportional to $k^{-3}$ \citep{2000A&A...355...17L}, estimated using IRIS at 100~$\mu$m on large angular scales in the "$k^{-3}$ regime" \citep{2007ApJ...665L..89L}, or considered to be negligible \citep{2009ApJ...707.1766V}. \citet{2011ApJ...737....2M} used the \citet{1998ApJ...500..525S} map at 100~$\mu$m as a cirrus template. This step was incorrect as, 
in the extragalactic fields, this map is highly contaminated (even dominated on some spatial scales) by the anisotropies of the CIB, as demonstrated in Sect. \ref{par:GC_100}\footnote{We therefore we did not compare their determination of the CIB with ours as their determination is highly biased owing to an incorrect cirrus removal.}.\\
The goal here is to examine the accuracy of the cirrus removal using \hi~data (as done by \citet{2011A&A...536A..18P}), with the ultimate aim of obtaining CIB anisotropy maps on fields larger than ELAIS N1, and even for a very large fraction of the sky with Planck data.
In this section, we present our power spectrum of the Galactic component using only 100~$\mu$m data on large scales. We then characterize the dust properties in our field. This is used in Sect. 5 to qualify the accuracy of the cirrus removal using \hi~data.
 
\subsection{Constraints on the cirrus contribution using IRIS 100 $\mu$m data} \label{par:IRIS}
\begin{figure}[!h]
\includegraphics[scale=0.5]{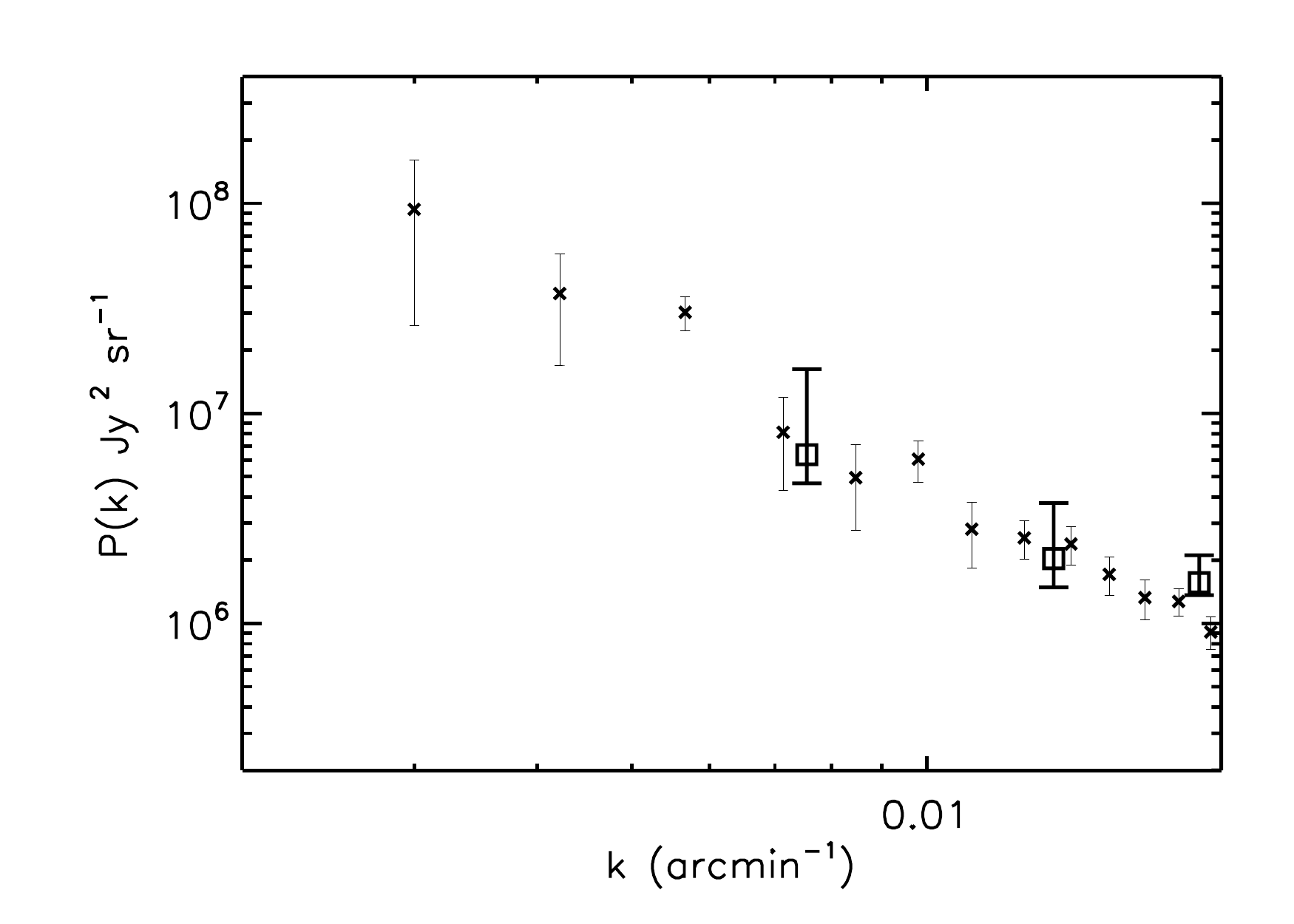}
\caption{Comparison of the cirrus power spectra computed with the IRIS 100 $\mu$m map on large scales with that computed with the 160 $\mu$m map. Crosses~: cirrus power spectrum at 100 $\mu$m computed on a 225 deg$^2$ 100~$\mu$m map centered on ELAIS N1 rescaled by the ratio $B_{100, cirrus}^{10 deg^2}/B_{100, cirrus}^{225 deg^2} \times (B_{160}/B_{100})^2$ 
is shown by crosses. Squares : power spectrum of ELAIS N1 computed at 160~$\mu$m.}
 \label{fig:pk_mips_iris_sc}
\end{figure}

The power spectrum of the Galactic component is first assumed to be a power law, as in \cite{2007A&A...469..595M}
\begin{equation}
P_{cirrus}(k)=P_0\left(\frac{k}{k_0}\right)^{\beta}, 
\end{equation}
where $P_0$ is the normalization of the power spectrum at $k_0=~0.01$~arcmin$^{-1}$. The cirrus component dominates the power spectrum on large scales ($k<0.01$~arcmin$^{-1}$). The ten square degree field selected at 160 $\mu$m does not allow us to probe this regime reliably, therefore a larger map is needed. After removing the sources, we fit the power spectrum of a 225 deg$^2$ IRIS 100~$\mu$m map centered on ELAIS N1 deriving $\beta = -2.53\pm0.03$ and $P_0=(4.93\pm0.20)\times 10^6$ Jy$^2$/sr. 
To compare this value to previous work (e.g. \cite{2007ApJ...665L..89L} in the Lockman Hole field), we need to rescale $P_0$ by the ratio of the cirrus brightness. We note that $P_0$ is indeed proportional to the square of the surface brightness of the cirrus $B_{cirrus}^2 =(B_{100}-B_{CIB})^2$. If we take $B_{CIB}= 0.78~$MJy/sr according to \citet{2000A&A...354..247L}, we obtain $B_{cirrus}^{Us} =1.25$~MJy/sr. Using $B_{cirrus}^{Lagache}=0.51$ MJy/sr, we then have
\begin{equation}\label{eq:norma}
P_{0, normalised}^{Lagache}=P_{0}^{Lagache}\times\left(\frac{B_{cirrus}^{Us}}{B_{cirrus}^{Lagache}} \right)^2=4.20 \pm0.93\times 10^6 \mbox{Jy$^2$/sr}
\end{equation}
which agree with our measured $P_{0}$.\\
The 100~$\mu$m power spectrum needs to be scaled by the ratio $(B_{160}$/$B_{100})^2$ in order to be compared directly with the 160 $\mu$m power spectrum \citep{2010ApJ...708.1611R}. This ratio depends on the cirrus physical properties and has thus to be determined sepcifically for our field. Unfortunately, this ratio is also scale-dependent: on large spatial scales, the brightness of the map is dominated by the cirrus, whereas on small spatial scales, CIB anisotropies dominate. We therefore differentiate the spatial scales larger than 95 arcmin from those smaller than 95~arcmin using a wavelet decomposition\footnote{IDL \textit{atrou} algorithm}. Moreover, we remove scales smaller than 6~arcmin as they are dominated by instrumental noise. We then perform a linear regression between the maps at 100 and 160 $\mu$m for both spatial scales, yielding the ratio $B_{100}/B_{160}$. To increase the statistics, we do a similar analysis in the Lockman Hole \citep{1986ApJ...302..432L}, using the data published by \citet{2007ApJ...665L..89L}. Our results are listed in Table \ref{tab:colors}. 
They are compared with previous large-scale cirrus $B_{100}/B_{160}$ color measurements \citep{1996AA...312..256B} and CIB colors from the \citet{2003MNRAS.338..555L} empirical model of galaxy evolution. On scales  $<$95~arcmin, our results are consistent with the CIB prediction of \citet{2003MNRAS.338..555L}. However, there is a discrepancy on scales  $>$ 95 arcmin between our results and those of  \citet{1996AA...312..256B}, which we attribute to real changes in cirrus properties from one field to another. \citet{2009ApJ...701.1450F} found that $B_{100}/B_{160}=0.25\pm0.01$ across the whole Taurus complex, whereas locally, they found that the same ratio varies from 0.27 to 0.5. \citet{2009ApJ...695..469B} computed colors in several small regions of the sky surrounding nearby galaxies ($\sim$0.1 deg), and they also found varying colors from one field to another: from $B_{100}/B_{160}$ = 0.36 to $B_{100}/B_{160}$=0.60. In both cases, they explained these differences by a variation of the interstellar radiation field and/or the abundance of very small and big grains. \\
To compare the cirrus power spectrum obtained using IRIS data at 100~$\mu$m with our 160~$\mu$m power spectrum, we rescale the 100~$\mu$m power spectrum by $B_{100, cirrus}^{10 deg^2}/B_{100, cirrus}^{225 deg^2}$ and then by $(B_{160}/B_{100})^2 =  (1/0.35)^2$, the large-scale color that corresponds to the cirrus color in our field. We show this comparison in Fig.~\ref{fig:pk_mips_iris_sc} and observe that they are in good agreement  for $k<$0.02~arcmin$^{-1}$ (i.e. in the "cirrus regime"). Unfortunately, the statistics of the large scales at 160~$\mu$m are quite poor and do not allow us to quantify the quality of the argument. \\
\begin{table*}\centering
 \begin{tabular}{*{4}{c}}
 \hline\hline
Field & $B_{100}/B_{160}$ & $B_{100}/B_{160}$ $>$95'& $B_{100}/B_{160}$ $<$ 95'\\
\hline
Lockman Hole & 0.62$\pm$0.01 & 0.30$\pm$0.01 & 0.64$\pm$0.01 \\
ELAIS N1 & 0.50$\pm$0.01 & 0.35$\pm$0.01 & 0.76$\pm$0.01 \\
\hline
Cirrus color &  & & \\
\cite{1996AA...312..256B}    & & 0.50      &\\
\citet{2009ApJ...701.1450F}  & & 0.27-0.5  &\\
 \citet{2009ApJ...695..469B} & & 0.36-0.6  & \\
\hline
CIB color &  &  &  \\
\cite{2003MNRAS.338..555L} & & &0.65 \\
\hline
\end{tabular}
\caption{Brightness ratios $B_{100}/B_{160}$ at scales larger and smaller than 95 arcmin in the Lockman Hole and in the ELAIS N1 field. The last four lines give the cirrus and CIB colors according to \cite{1996AA...312..256B}, \citet{2009ApJ...701.1450F}, \citet{2009ApJ...695..469B}, and \cite{2003MNRAS.338..555L}, respectively.}
\label{tab:colors}
\end{table*}

\subsection{Dust colors and dust-\hi~emissivities}\label{par:dust+hi}
The dust that is heated by the interstellar radiation field and emits in the IR is mixed with neutral hydrogen. Thus, the IR emission of the cirrus is strongly correlated with the \hi~21 cm line. \citet{1988ApJ...330..964B} showed that this correlation is tight at high Galactic latitudes at 60 and 100 $\mu$m. This correlation has often been used to study dust properties, for instance by \citet{1996AA...312..256B} who derived the dust spectrum associated with \hi~gas.\\
In this section, we use GBT data at 21 cm to derive the FIR emission of the cirrus, which is then removed from our data at 100~$\mu$m and 160~$\mu$m in Sect.~\ref{par:removal}.\\
In the ELAIS N1 field, there are three distinguishable \hi~velocity components : the local component, an intermediate velocity cloud (IVC), and a high velocity cloud (HVC). We first compute their integrated emission by adding all velocity channels with -14 km/s $<V_{LSR}<$ 43 km/s for the local, -79 km/s$<V_{LSR}<$-14 km/s for the IVC, and -163 km/s $<V_{LSR}<$-79 km/s for the HVC. Second, assuming the optically thin case, we estimate their column density using
\begin{equation}
N_{HI}(x,y)=1.823\times 10^{18} \sum_v T_B(x,y,v)dv,
\end{equation}
where $N_{HI}$ is the \hi~column density in unit of 10$^{20}$ atoms/cm$^2$, $T_b$ the brightness temperature, and $v$ the velocity.\\
The interstellar medium in the ELAIS N1 field seems to be dominated by
neutral atomic hydrogen that reaches a peak N$_{HI} \approx 1.5
\times 10^{20}$ cm$^{-2}$ in each of the three components (Fig. \ref{fig:GBT}).
The brightness temperature of the \hi~line is always $\leq 8.9$~K. 
Since molecular hydrogen, $H_2$, begins to be seen in
directions where $N_{HI} > 2 \times 10^{20}$ cm$^{-2}$
and $T_b > 12$ K (e.g., \citet{2006ApJ...636..891G}, \citet{2002A&A...389..393L}, \cite{2005AJ....129.1968L}), it is unlikely that there are significant amounts of $H_2$ in our field. Therefore, we can apply the decomposition following \cite{2005ApJ...631L..57M}
\begin{equation}
I_{\lambda}=\sum \alpha_{\lambda}^i N_{HI}^i(x,y) + C_{\lambda}(x,y),
\label{eq:mod}
\end{equation}
where $I_{\lambda}$ is the infrared map, $N_{HI}^i(x,y)$ is the column density of the $i$-the \hi~component, $\alpha_\lambda^i$ is the emissivity of component $i$ at wavelength $\lambda$, and $C_{\lambda}(x,y)$ is a residual term (offset + CIB). 
The correlation coefficients $\alpha_\lambda^i$ are estimated using a $\chi^2$ minimization\footnote{we use the IDL function \textit{regress}}. The error bars given by the IDL function are valid only if the noise of the $I_{\lambda}$ maps is Gaussian and if the noise affecting N$_{HI}$  is negligible. This may not be the case as the maps contain the IRIS or MIPS instrumental noise and CIB anisotropies. \citet{2011A&A...536A..24P} carried out Monte Carlo simulations to estimate the errors in $\alpha_\lambda^i$ for IRIS 100 and 60~$\mu$m and determined the coefficients by which they multiplied the error bars found assuming a Gaussian noise. We multiply our errors by these coefficients at 60 and 100~$\mu$m. For MIPS at 160~$\mu$m, we take the mean of the 100 and 350~$\mu$m coefficients determined by \citet{2011A&A...536A..24P}, as these coefficients vary only slightly with wavelength. They are on the order of eight.\\
The emissivities $\alpha^i_{\lambda}$ are computed at 60, 100, and 160~$\mu$m in the ELAIS N1/MIPS field (i.e. ELAIS N1 field restricted to the MIPS coverage) and at 60 and 100~$\mu$m only in the ELAIS N1/GBT field (i.e. the entire field covered by \hi~data, see Fig. \ref{fig:GBT}). Fig. \ref{fig:GBT} shows that ELAIS N1/MIPS does not contain the IVC. Therefore, for this field, we use only two components, the local and the HVC, to avoid any additional noise. Our results are given in Table~\ref{tab:only_n1}.\\
The emissivities of the local component are in accordance with those found by \citet{2005ApJ...631L..57M} at the three wavelengths. They used \hi~observations from the GBT \citep{2005AJ....129.1968L} to compute emissivities at 24, 60, 100, and 160 $\mu$m in the Spitzer Extragalactic First Look Survey field. They have two IVCs whose emissivities are in agreement with ours at 100 $\mu$m. However, there is a discrepancy at 60~$\mu$m. For IVC1 and IVC2, \citet{2005ApJ...631L..57M} found $B_{60}/B_{100}$= 0.50 and 0.34 respectively, whereas we have $B_{60}/B_{100}$=0.30. This value is in line with \citet{2011A&A...536A..24P}, who found 0.23$<B_{60}/B_{100}<$0.42 in 14 fields covering 800 square degrees of the sky. They interpreted these differences as a contamination of the emission at 60 $\mu$m by non-equilibrium emission due to a higher relative abundance of very small grains compared to big grains at shorter wavelength. \\
\begin{figure*}
\includegraphics[scale=0.33]{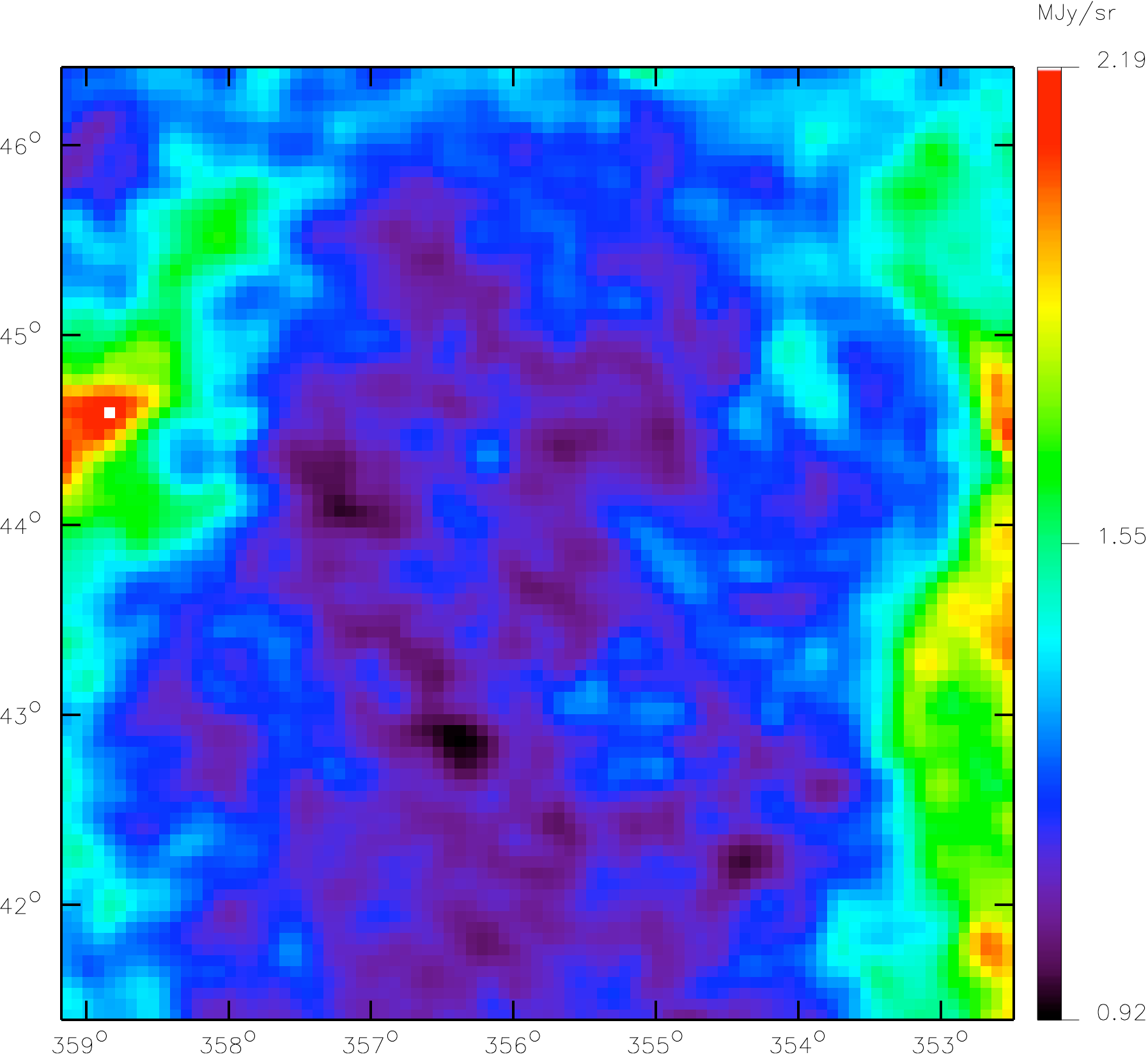}\includegraphics[scale=0.33]{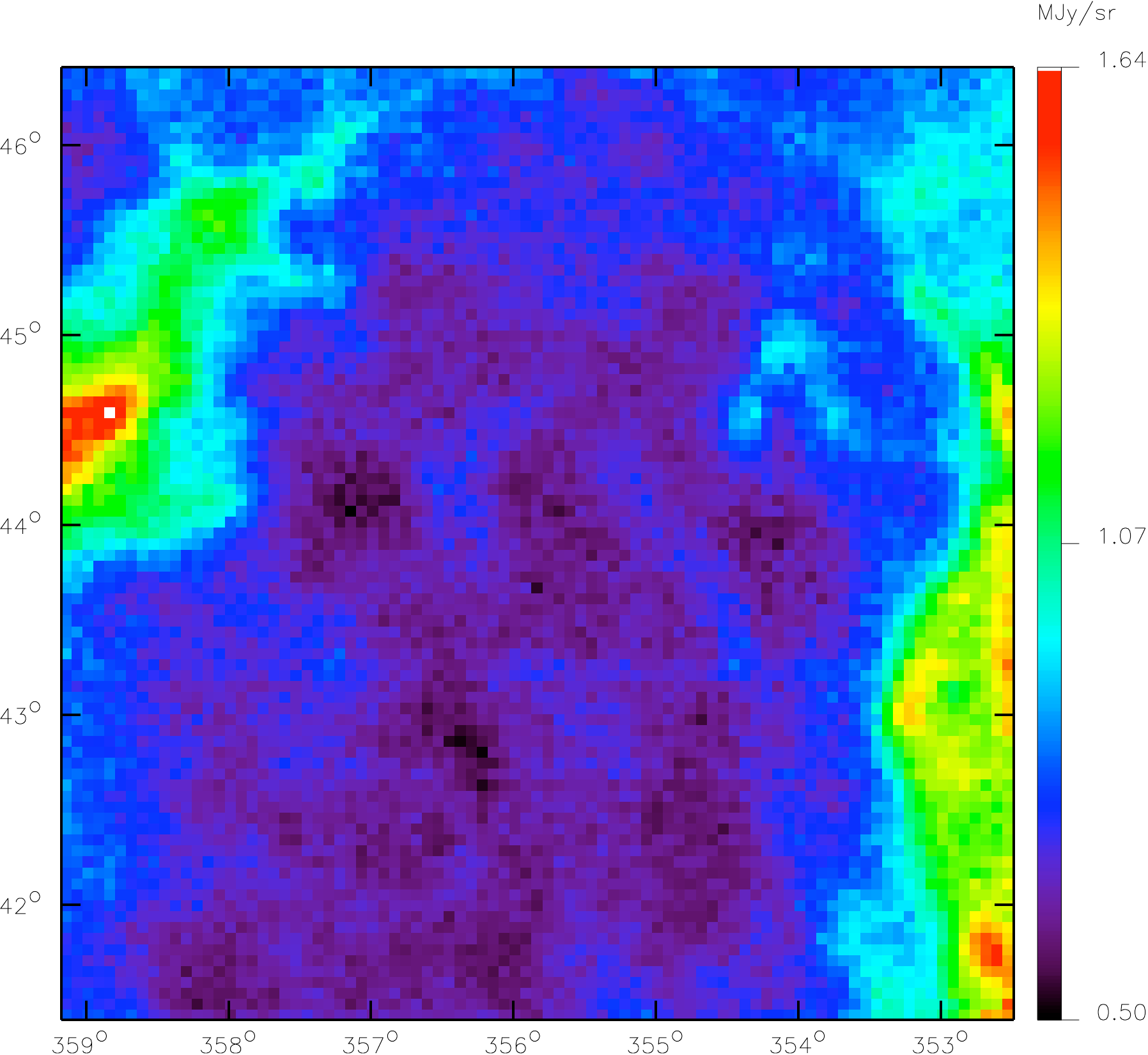}\includegraphics[scale=0.33]{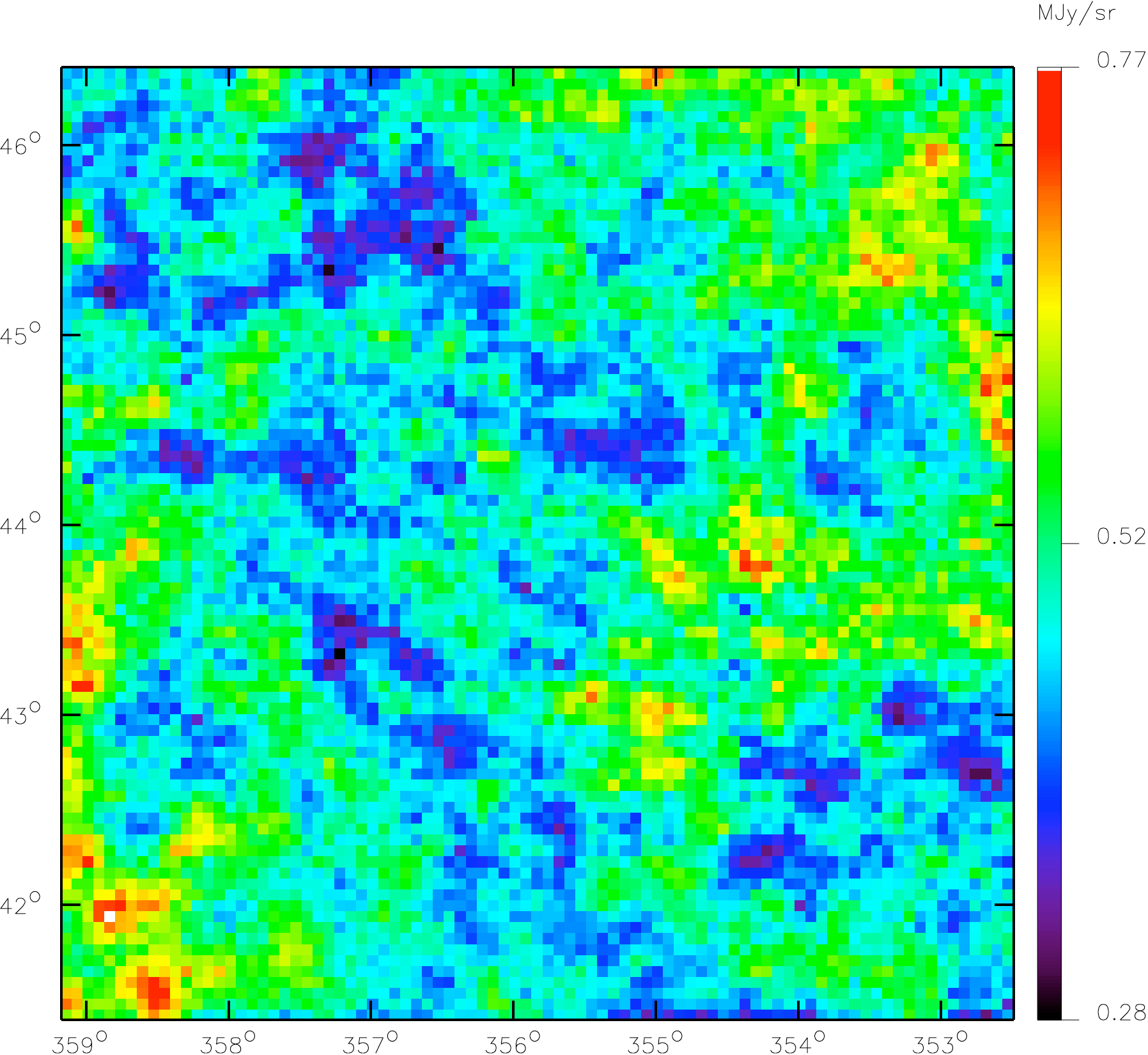}
\caption{The left panel shows the IRIS 100 $\mu$m map projected on the GBT map and convolved with the GBT beam. The emission of the cirrus (local+IVC+HVC) at 100~$\mu$m (the emissivity $\times N_{HI}$) is shown in the middle panel and the right one represents the IRIS 100~$\mu$m map \textbf{convolved by the beam of the GBT} from which we have subtracted the cirrus emission.}
\label{fig:iris_ms_cirrus}
\end{figure*}
Moreover, the HVC in ELAIS N1 is detected at 160~$\mu$m. This confirms the detection of dust in some HVCs.
Finally, we note that our results are also in very good agreement with those from \citet{2011A&A...536A..24P}, who performed a similar analysis on the same field.\\
Using emissivities, colors can be computed and compared to previous works. We find that $B_{160}/B_{100}$ = 2.87 for the local component. \citet{2009ApJ...695..469B} removed the mean value of the CIB, 0.78 MJy/sr from their $<B_{100}>$, and they observed an increase in the ratio $B_{160}/B_{100}$ with $<B_{100}>$ meaning that brighter regions are colder. We subtract the same CIB level of 0.78 MJy/sr from our $<B_{100}>$ , even though we get a mean CIB residual of 0.58 MJy/sr (see Tab \ref{tab:only_n1}). This results in $<B_{100}>$= 0.4 MJy/sr, which is much lower than any of their $<B_{100}>$. They kept only regions with $<B_{100}> >$2.5 MJy/sr, in order to be dominated by variations in cirrus emission (and not be contaminated by CIB anisotropies). Their fields have a typical $B_{160}/B_{100}$ of 2, which is very close to our value. Our much lower $<B_{100}>$ implies that $B_{160}/B_{100}$ may reach a plateau at a value of $\sim2$~MJy/sr for $<B_{100}>$ lower than 2.5 MJy/sr. The value of $B_{160}/B_{100}$ can also be compared to those given in Table~\ref{tab:colors}, which were computed with a linear regression between the two maps at large scales. We achieve an excellent agreement, finding $B_{100}/B_{160}$ = 0.35$\pm$0.01 in the two cases.\\
The CIB should be the only astrophysical component contained in the residual map. To test this, we can check whether the residual mean value agrees with the mean value of the CIB. This is the case at 160~$\mu$m, after the map is corrected by an offset term (see Sect.~\ref{par:tpm}). At 100~$\mu$m, there is a discrepancy between the prediction and our result, which we attribute to residuals of zodiacal light in the map. The IRIS data were indeed calibrated in the DIRBE data, which contain a residual zodiacal emission that leads to an overestimate of the CIB (see \cite{2006A&A...451..417D}). At 60~$\mu$m, we do not try to measure the mean value of the residual because the CIB level is on the order of the residual of the zodiacal light at this wavelength \citep{2001A&A...371..771R}.
\begin{table*}
 \begin{tabular}{*{8}{c c c| c c c|c c }}
  \hline\hline
&&&\multicolumn{3}{|c|}{Dust/\hi~correlation coef in ELAIS N1/MIPS}&\multicolumn{2}{c}{Dust/\hi~correlation coef in ELAIS N1/GBT}\\
						\hline
						
$\lambda$               	&$(\mu$m)&      	&60 		        &100			&160			&60			&100\\
$\alpha^{local}_{\lambda}$	&	 &this paper	&0.175 $\pm$  0.056	&0.87 $\pm$ 0.14 	&2.46 $\pm$ 0.43	&0.175 $\pm$ 0.014      &0.877 $\pm$ 0.028 \\
$\alpha^{IVC}_{\lambda}$  	&	 &this paper	&		        &			&			&0.207 $\pm$ 0.015	&0.699 $\pm$ 0.038\\
$\alpha^{HVC}_{\lambda}$ 	        &  	 &this paper	&-0.004 $\pm$ 0.015	&0.034 $\pm$ 0.031	& 0.31 $\pm$ 0.08 	&-0.001 $\pm$ 0.007	&0.010 $\pm$ 0.023\\
\hline
$\alpha^{local}_{\lambda}$	&        & MAMD 2005    &0.16 $\pm$ 0.02 	&0.80 $\pm$ 0.08	&1.7 $\pm$ 0.02	         &			&\\
$\alpha^{IVC1}_{\lambda}$  	&	 & MAMD 2005	&0.35 $\pm$0.04		&0.70 $\pm$0.09	&2.7 $\pm$0.4		&&\\
$\alpha^{IVC2}_{\lambda}$  	&	 & MAMD 2005	&0.31$\pm$0.04		&0.9$\pm$0.1		&1.4 $\pm$0.4		&&\\
$\alpha^{HVC}_{\lambda}$  	&	 & MAMD 2005	&0.05$\pm$0.01		&0.055$\pm$0.015	&0.8$\pm$0.1		&&\\
\hline
$\alpha^{local}_{\lambda}$	&	 & Planck 2011	&  			&		        &			&0.166$\pm$0.011  	&0.862$\pm$0.033\\
$\alpha^{IVC}_{\lambda}$  	&	 & Planck 2011	&			&			&			&0.213$\pm$0.012	&0.723$\pm$0.036\\
$\alpha^{HVC}_{\lambda}$ 	        &        & Planck 2011	&		        &			&			&-0.001$\pm$0.007	&-0.009$\pm$0.022\\
\hline
$<residue>$ 			&(MJy/sr)&this paper	&			& 0.58			&0.76\tablefootmark{a}	&			&\\
CIB mean		        	&(MJy/sr)&B\'ethermin 2011&			& 0.30$\pm$0.01		&0.63$\pm$0.01		&					&\\
\hline\hline
\end{tabular}

\caption{Emissivities and CIB levels. Columns 4, 5, 6 : Emissivities (in units of MJy/sr (10$^{20}$ H atoms)$^{-1}$ cm$^2$) computed in ELAIS N1/MIPS at 60, 100 and 160 $\mu$m compared to those of the literature. Columns 7, 8 : Emissivities computed in N1/GBT at 60, 100 $\mu$m. Uncertainties are 1 $\sigma$ uncertainties that take into account the statistical variance and the instrumental noise. MAMD 2005 stands for \citet{2005ApJ...631L..57M}. Their emissivities are for the Spitzer XFLS field. Planck 2011 stands for \citet{2011A&A...536A..24P}. Their emissivities are for the ELAIS N1/GBT field. The second last line gives the mean values of the residual maps. At 160~$\mu$m, the mean value has been corrected for the offset determined in Sect. 6.1, and gives the level of the CIB (although see Sect. 6 for a more accurate determination). At 100~$\mu$m, the mean value is strongly contaminated by residual zodical emission (see \cite{2006A&A...451..417D}) thus does not give the CIB level. The last line gives the CIB coming from the \cite{2011A&A...529A...4B} model.
\tablefoottext{a}{This value has been corrected from an offset that is present in the scanning map (see Sect.~\ref{par:tpm}).}}
\label{tab:only_n1}
\end{table*}
\subsection{Dust temperatures}
The IR/\hi~emissivities give constraints on the dust temperature.  We assume that the emission of big grains at thermal equilibrium with a radiation field is a modified black body
\begin{equation}
I_{\nu}=\tau_{\nu}B_{\nu}(T_{BG}),
\end{equation}
where $B_{\nu}$ is the Planck function, $T_{BG}$ the big grains' equilibrium temperature, and $\tau_{\nu}$ the optical depth. It can be expressed as $\tau_{\nu}=N_{HI}\epsilon_{\nu}$ with $\epsilon_{\nu}$ the dust emissivity per H atom that is usually assumed to be a power law $\epsilon_{\nu}= \epsilon_0(\nu/\nu_0)^{\beta}$ where $\beta$ is the emissivity spectral index. Following \citet{1996AA...312..256B}, we assume that $\beta$ = 2. At wavelengths where the emission is dominated by big grains, the IR/\hi~correlation coefficients can be written as $\alpha_{\nu}=\epsilon_{\nu}B_{\nu}(T_{BG})$. We use the infrared-\hi~correlation coefficients at 100 and 160 $\mu$m to estimate the big grains temperature $T^i_{BG}$ of each \hi~component assuming that the contribution of very small grains in the far-infrared is negligible.
\begin{equation}
\frac{\alpha_{100}^i}{\alpha_{160}^i}=\frac{B_{\nu}(100 \mu m,T^i_{BG})}{B_{\nu}(160 \mu m,T^i_{BG})}\left(\frac{160}{100}\right)^2.
\end{equation}
We find $T_{BG}^{local} = 15.9\pm0.2$ K. This agrees with \citet{2009ApJ...695..469B}, who found dust temperatures in the diffuse medium of between 15.7 K and 18.9 K and with \cite{2005ApJ...631L..57M} who found $T_{BG}^{local}$ between 16.3 and 18.8K. 
We do not detect the HVC at 100 $\mu$m, hence we cannot determine its temperature. However, we compute a limit to its temperature assuming a 3$\sigma$ limit for the detection at 100 $\mu$m. We get $T_{BG}^{HVC} < 15$K, which agrees with \citet{2005ApJ...631L..57M} who found a value of $T_{BG}^{HVC}$ between 9.9 K and 11.6 K. The HVC is colder than the local diffuse medium. This is consistent with a lower radiation field than in the solar neighbourhood owing to the distance of the HVC.

\section{Removal of the Galactic component}\label{par:removal}
\subsection{Removal of the cirrus either spatially or by considering the power spectra and error bars}\label{par:eb2}
The cirrus contribution removal can be done in two ways: spatially in the maps or by subtracting the cirrus power spectrum from that of the infrared map. We first need to know whether these two methods are equivalent. We consider the power spectrum of the CIB, $\hat{P}_{CIB}(k)$, as an estimator for this test, and compute its variance for the two methods in a naive approximation of Gaussianity. We have two maps A and B, containing, respectively, CIB + Galactic component (GC) and only the GC. In Fourier space, we have $a_k^A = a_{k,CIB}+a_{k,GC}$ and $a_k^B = a_{k,GC}$, where $a_k$ are the Fourier coefficients. The power spectrum can be written 
\begin{equation}
\hat{P}_{CIB}(k) = <a_{k,CIB}^2>.
\end{equation}
The mean values of the power spectrum of the CIB anisotropies are equal for both methods but not their variances. For the spatial removal, 
\begin{eqnarray}
Var(\hat{P}_{CIB}) &=& <(a_k^A-a_k^B)^2> - <(a_k^A-a_k^B)>^2\\
	                        &=& Var(P_{CIB}), 
\end{eqnarray}
whereas for the removal at the power spectrum level a correlation term between the cirrus power spectrum and that of the CIB appears, in addition in the variance of the CIB power spectrum
\begin{eqnarray}
Var(\hat{P}_{CIB}) &=& <(a_k^Aa_k^A-a_k^Ba_k^B)^2>-<(a_k^Aa_k^A-a_k^Ba_k^B)>^2 \\
                   	      &= &Var(P_{CIB}) + Var(P_{CIB})\times (a_k^B)^2.\\
\end{eqnarray}
Thus, from this simple argument we expect error bars to be smaller after the spatial subtraction. We check this by performing simulations using mock data. We generate signal plus noise maps and analyze them in both ways. Error bars are indeed smaller in the spatial subtraction case. The template (spatial) subtraction removes each moment of the statistics, whereas the second method only removes the moment of the power spectrum. 

\begin{figure}
\includegraphics[scale=0.5]{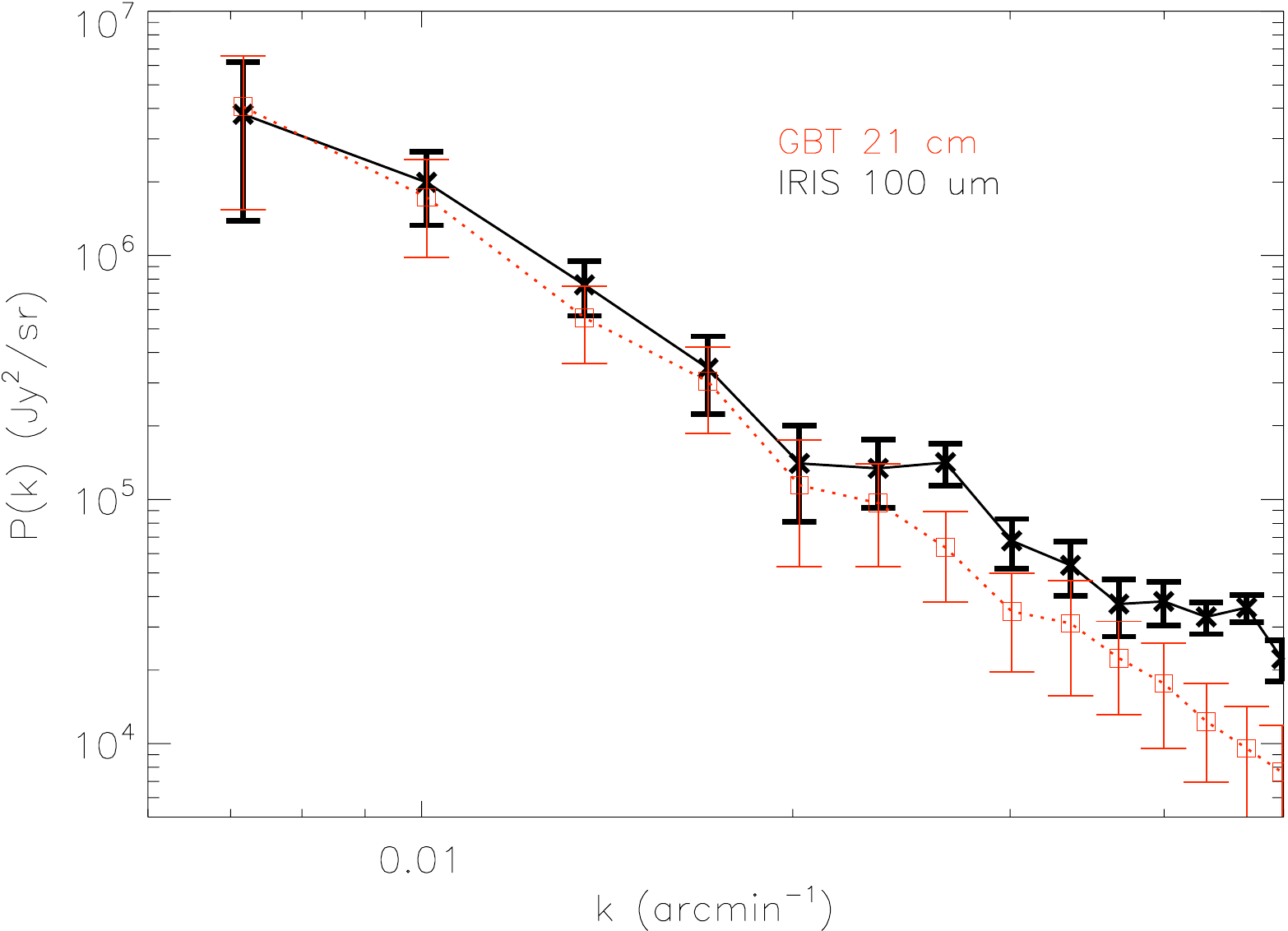}\centering
\caption{The black continuous line with the crosses and bold error bars show the power spectrum of IRIS 100 $\mu$m and the red dotted line with squares and thin error bars represents the power spectrum of the emission of the cirrus at 100 $\mu$m calculated from the \hi~map. The discrepancy observed for k$>$0.02 arcmin$^{-1}$ is due to the clustering of SB galaxies (see Sect.~\ref{par:CIB100}).}
 \label{fig:pk_iris_gbt}
\end{figure}

\begin{figure}[h]
\begin{center}
\includegraphics[scale=0.48]{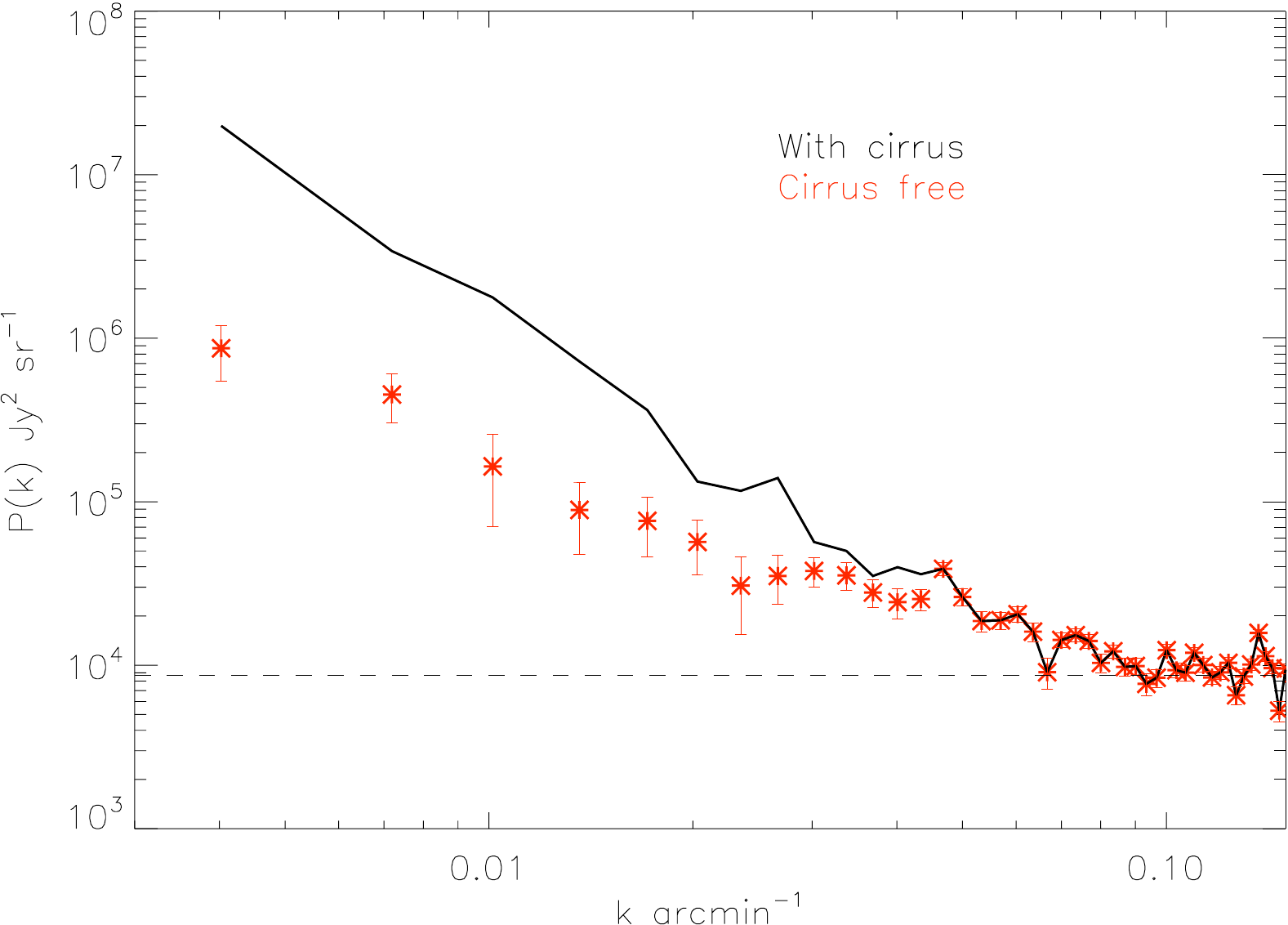}\centering
\caption{Red crosses show the CIB anisotropies power spectrum computed at 100~$\mu$m. The contribution of the cirrus has been removed following Sect. \ref{par:GC_100}. The power spectrum of the noise has been subtracted and the resulting power spectrum has been divided by the power spectrum of the PSF. The black line is the power spectrum of the map that contains the cirrus contamination. The dashed line shows the level of the Poisson noise which is 8.7$\times 10^3$Jy$^2$/sr.}
\label{fig:pk_cib_fin}
\end{center}
\end{figure}

\subsection{Contribution of the Galactic component to the power spectrum at 100 $\mu$m}\label{par:GC_100}
We start by comparing the cirrus power spectrum obtained at 100~$\mu$m to that computed with \hi~data in the ELAIS N1/GBT field. To construct a map of the cirrus emission at 100~$\mu$m, we take the sum of each component column density weighted by the emissivities determined in Sect. \ref{par:dust+hi}~:
\begin{equation}
B_{100} = \alpha^{local}_{100}\times N^{local}_{HI}+\alpha^{IVC}_{100}\times N^{IVC}_{HI}+\alpha^{HVC}_{100}\times N^{HVC}_{HI}.
\end{equation}
The left panel of Fig. \ref{fig:iris_ms_cirrus} shows the map at 100~$\mu$m projected onto the GBT map and convolved with the GBT beam, the middle panel shows $B_{100}$ computed from 21cm data. The right panel shows the 100 $\mu$m cirrus-free residual. We can clearly see that most of the contribution of the cirrus is removed and that the residual map contains only the CIB. This is even more visible in Fig. \ref{fig:pk_iris_gbt}, which compares the 100~$\mu$m power spectrum to that of $B_{100}$ computed from 21 cm data. They are in very good agreement for $k<$0.02~arcmin$^{-1}$. The discrepancy observed on smaller scales (k$>$0.02 arcmin$^{-1}$) is expected, as the clustering of SB galaxies dominates (see Sect \ref{par:CIB100}). Therefore, this shows that using of IR/\hi~emissivities is a reliable method for measuring the contribution of the Galactic component in the far-infrared.\\

We can now subtract this cirrus emission from the original map to get the CIB anisotropy map. Since the FWHM of the GBT is much larger than that of IRIS, the 100~$\mu$m ELAIS N1/GBT map must be convolved with the GBT beam (9.1 arcmin). Since the cirrus contamination is primarily on large scales, we can create a hybrid power spectrum using the power spectrum of the CIB anisotropies obtained by removing the \hi~on large scales, and that of the original map on small  scales. We use $k=0.05$~arcmin$^{-1}$ for the transition between the two power spectra. We note that the power spectrum of the CIB anisotropies on large scales must be divided by the power spectrum of a PSF with a FWHM = $\sqrt{(FWHM_{GBT}^2-FWHM_{IRIS}^2)}$ to take into account the convolution by a larger beam. The noise map is computed following Sect. \ref{par:pk_and_eb} and the PSF is that of Sect. \ref{par:data_hi_IRIS}. We subtract the noise power spectrum from the raw one and divide it by the power spectrum of the PSF. We present the power spectrum as well as the cirrus-free one on Fig. \ref{fig:pk_cib_fin}. We clearly see the amount of power due to the cirrus that has been removed. We derive a Poisson noise of 8690$\pm190$ Jy$^2$/sr, which disagrees with \citet{2002A&A...393..749M} who found $\sim$5.8$\times 10^3$Jy$^2$/sr, although the bright point sources are removed in the same way. The main difference between both studies is the removal of the cirrus contribution. They fitted the large-scale part of the 100 $\mu$m power spectrum and removed the fit on all scales. When we carry out the same analysis, we find a Poisson noise $\sim$ 5$\times 10^3$Jy$^2$/sr, in agreement with them. We conclude that they overestimated the cirrus contribution by removing a power law on all scales applying a fit only at large scales. The power-law fit is indeed contaminated by CIB anisotropies. 

\begin{figure}
\includegraphics[scale=0.47]{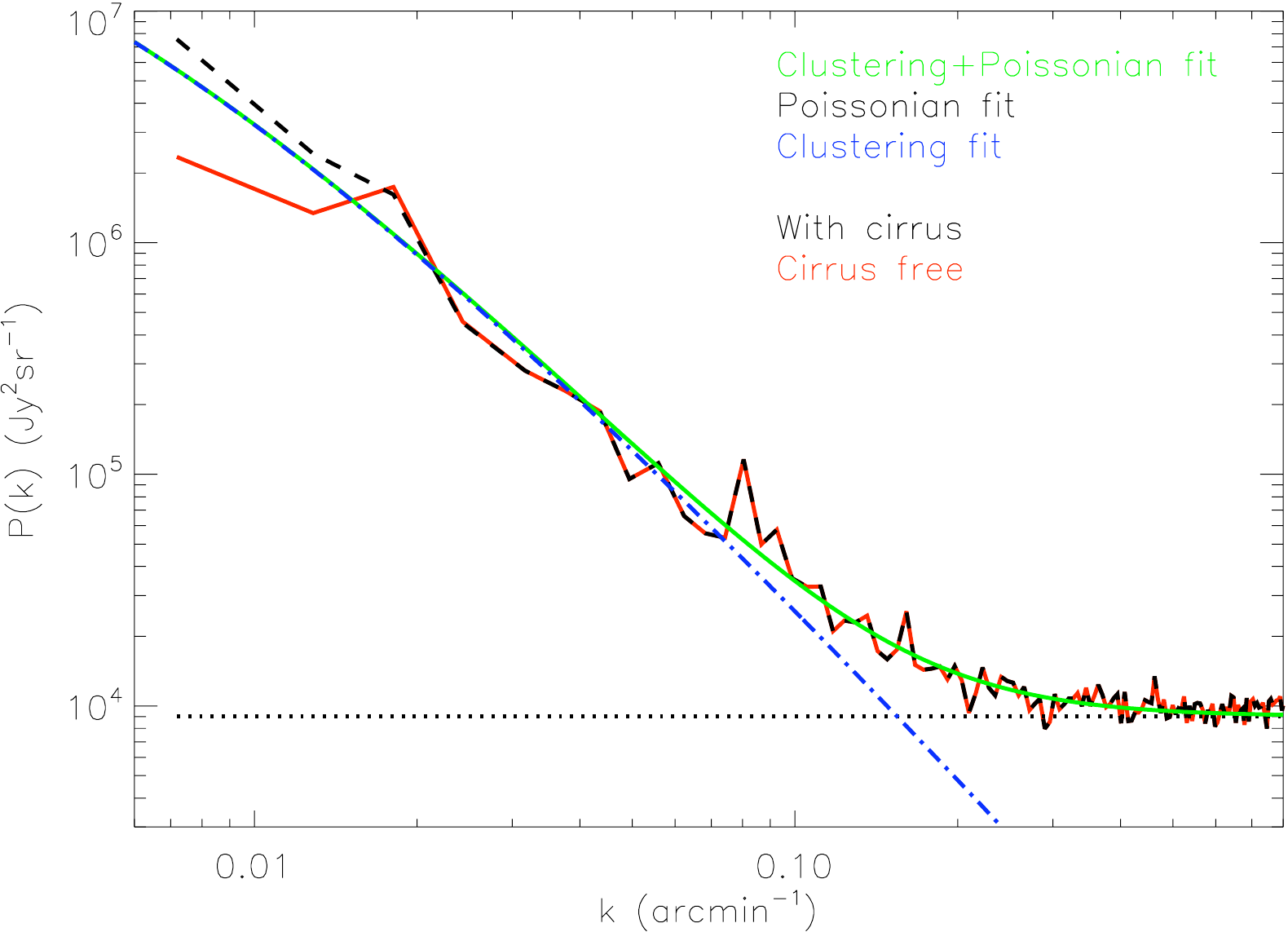}\centering
\caption{MIPS power spectrum at 160~$\mu$m. The dashed black line shows the power spectrum of the raw map (noise subtracted and divided by the PSF power spectrum). The red line represents the power spectrum of the map to which the cirrus emission has been subtracted. The blue dash-dotted line shows our fit to the power spectrum of the clustering and the black horizontal dotted line is the fit to the Poisson noise level. The green line shows the sum of the clustering and Poisson components. Error bars are not shown for display purposes.}
 \label{fig:pk_cib+cirrus}
\end{figure}

\begin{figure}[h]
\begin{center}
\includegraphics[scale=0.52]{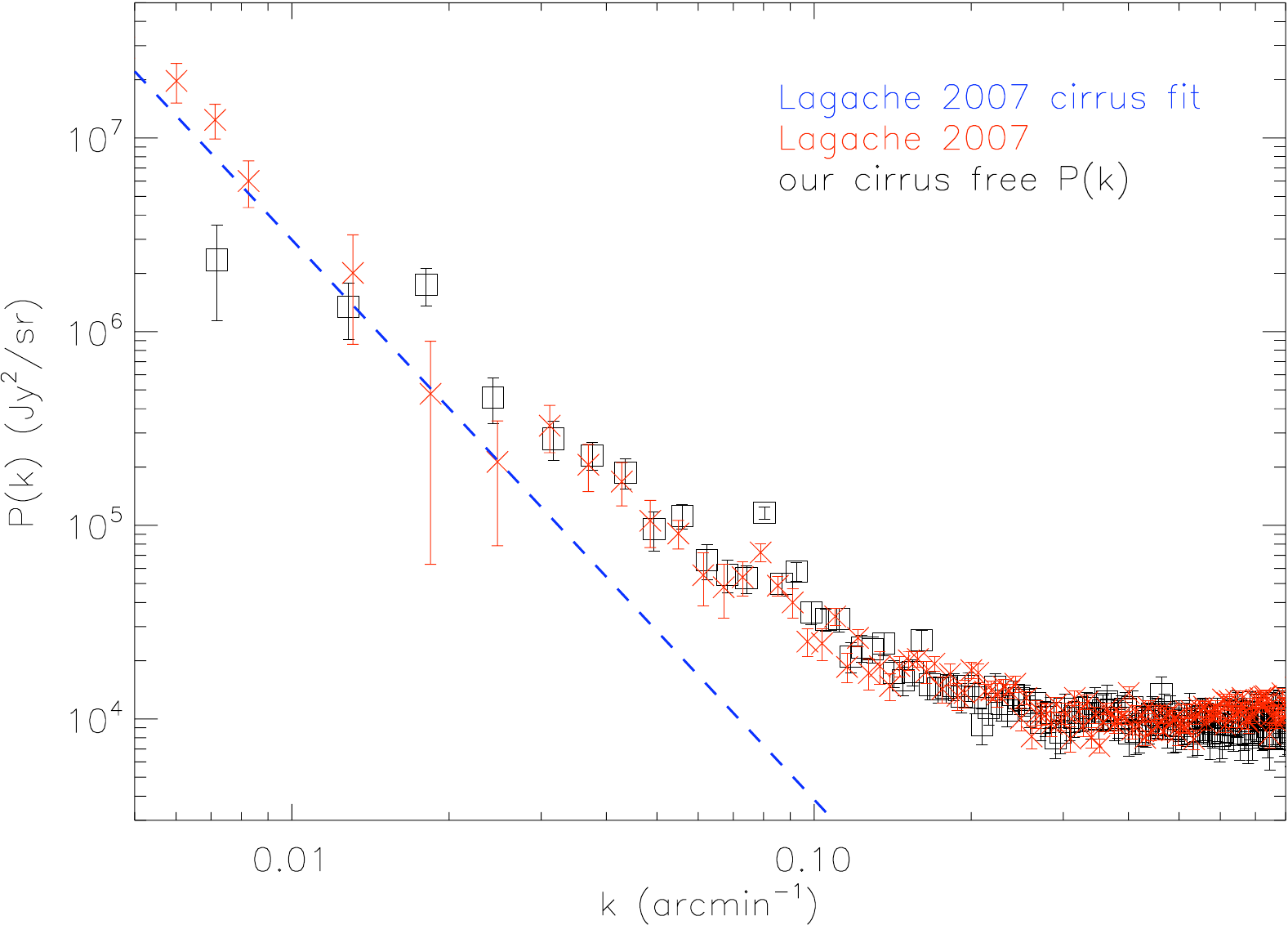}
\caption{Black squares are the resulting power spectrum of the CIB anisotropies computed by removing the cirrus contamination using \hi~data in the ELAIS N1 field. Red crosses represent the \citet{2007ApJ...665L..89L} power spectrum computed in the Lockman-Hole field. It includes both CIB and cirrus anisotropies and the blue dashed line shows their estimate of the cirrus power spectrum.} 
\label{fig:cib_fin}
\end{center}
\end{figure}

\subsection{Contribution of the Galactic component to the power spectrum at 160 $\mu$m}\label{par:GC_160um}
Using the same method as in Sect. \ref{par:GC_100}, we remove the cirrus emission from the ELAIS N1/MIPS map at 160~$\mu$m. We compute the hybrid power spectrum with a cut at $k=0.05$~arcmin$^{-1}$. Fig. \ref{fig:pk_cib+cirrus} shows the total power spectrum (black) and the cirrus-free one (red). We clearly see the difference only on the largest scales available with this map. We also plot our fit to the clustering power spectrum from \citet{2007ApJ...665L..89L} in blue, and the shot noise level in black. The green line shows the sum of the two-component fits. We also compare our resulting power spectrum to that of \citet{2007ApJ...665L..89L} in Fig. \ref{fig:cib_fin}. These power spectra are in very good agreement for $k>$ 0.03 arcmin$^{-1}$, where they are dominated by the CIB anisotropies (both clustering and Poisson noise). On scales $<$ 0.01 arcmin$^{-1}$, there is more power in the \citet{2007ApJ...665L..89L} power spectrum because it contains the cirrus contribution (the blue dashed line is their estimate of the power spectrum of the cirrus). We can see that using \hi~data, we are able to extend the measurement of the correlated fluctuations to large scales. This shows that making use of \hi~data at 21 cm is an efficient way of removing the contamination of the Galactic component.\\

\begin{figure*}[]
  \centering\includegraphics[scale=0.6]{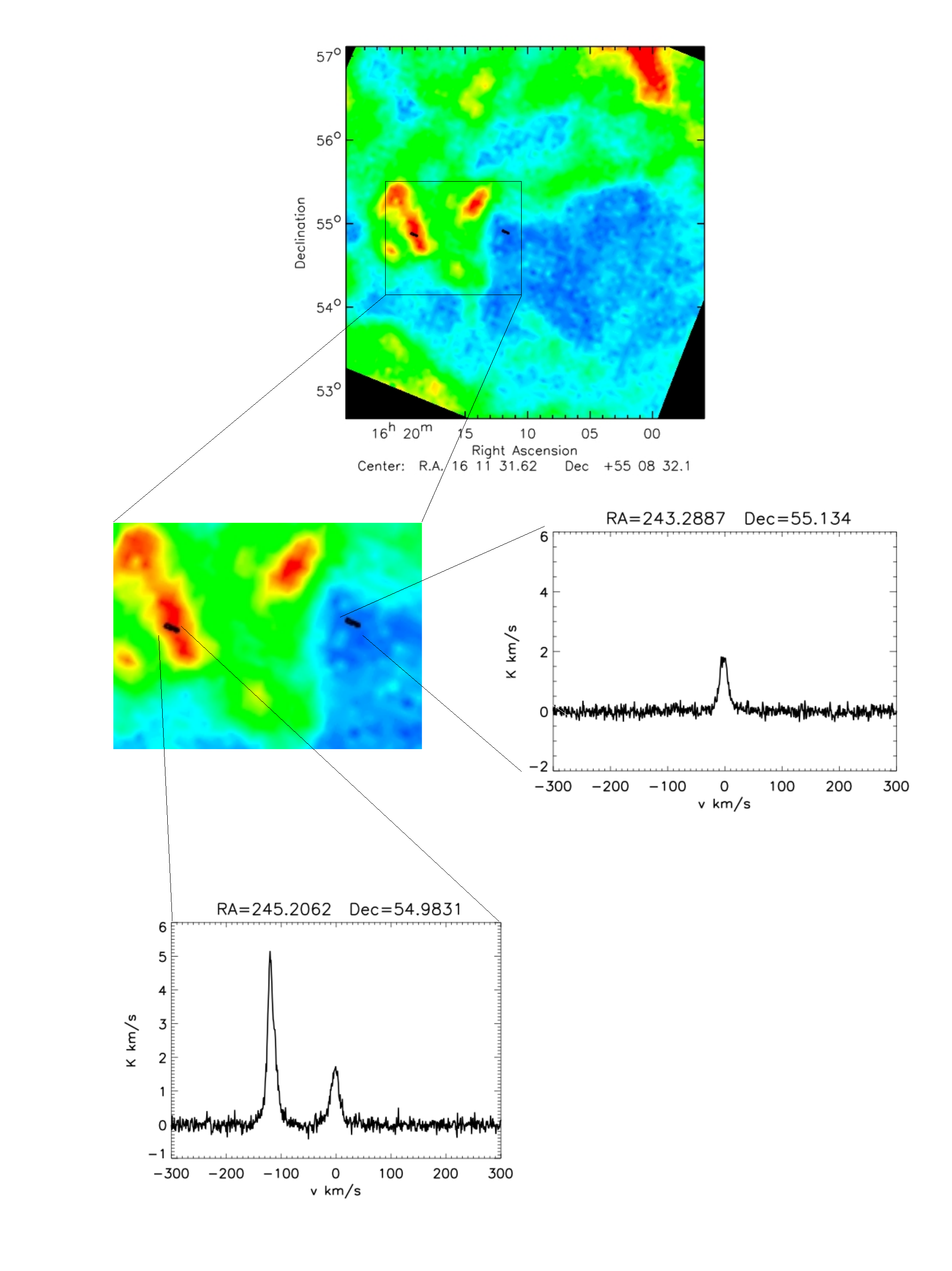}
  \caption{Positions of the total power mode observations on the GBT map with their \hi~velocity profiles.}
  \label{fig: tpm_sur_gbt}
\end{figure*}

\section{Measurement of the CIB mean levels at 100~$\mu$m and 160~$\mu$m}
\subsection{CIB mean estimate with Total Power Modes at 160~$\mu$m }\label{par:tpm}

We can also use our \hi~data and emissivity measurements to compute the absolute level of the CIB at 160~$\mu$m. We consider the two TPMs of {\it Spitzer} archival observations (26961920 \& 26962432) of the ELAIS N1 field that were designed to cross-check the calibration of the diffuse emission at 160~$\mu$m. Even with a cryogenic telescope such as {\it Spitzer}, there is a small component of thermal emission at longer wavelengths that contaminates the standard photometric observations. The TPM mode by-passes the effects of this spurious radiation by comparing the emission of the target (sky) with that of an internal dark to provide an absolute measurement (see MIPS Handbook $\S$ 3.1.12). This mode was designed precisely to observe relatively faint extended emission regions. The TPM observations that we used are discussed in Sect. \ref{par:tpmobs}.\\
There is an HVC in TPM 1 and only the local component in TPM 2 as shown on Fig. \ref{fig: tpm_sur_gbt}. We first compare the MIPS scan map values to the TPMs.
By first subtracting the map values from those of the TPMs (see Table \ref{tab:tpm}), we determine the offset of the scan map from each TPM position. They are in good agreement in the two regions with an average offset of -2.05$\pm$0.24 MJy/sr, which has no consequences for the power spectrum estimate as well as on the CIB level determination that follows. \\
Making use of the TPM values and the emissivities previously computed, we determine the absolute level of the CIB at 160 $\mu$m.
After subtracting the zodiacal light, the components of the TPM are
\begin{equation}
\label{eq_TPM}
TPM  - Zodiacal = Cirrus + CIB, 
\end{equation}
where the cirrus brightness is 
\begin{equation}
\label{eq_TPM_cirrus}
Cirrus = \alpha^{local}\times N_{HI}^{local} +  \alpha^{IVC}\times N_{HI}^{IVC} +  \alpha^{HVC}\times N_{HI}^{HVC}, 
\end{equation}
and $\alpha$ are the values computed across the whole ELAIS N1/MIPS field, as listed in Table \ref{tab:only_n1}. We use emissivities calculated over the entire field rather than over the TPM regions alone since they are far too small to obtain an accurate measurement.\\
The zodiacal light, which has a value of $B_{zodi}=0.83\pm$0.12 MJy/sr, was previously subtracted. It was estimated using the Spitzer background model (Reach 2000; \citet{1995Natur.374..521R}; \citet{1998ApJ...508...44K}). With our emissivities, we can compute the cirrus brightness of both TPMs (following Eq. \ref{eq_TPM_cirrus}) and thus compute the CIB level (following Eq. \ref{eq_TPM}). Our results are listed in Table \ref{tab:tpm}. The last column gives the CIB levels: they are in excellent agreement, even though they were computed in two regions with completely different cirrus contaminations. We get an average value of $B_{160}=0.77 \pm$0.04$\pm$0.12 MJy/sr for the CIB level at 160~$\mu$m. The first error is statistical, and the second is systematic and dominated by the error in the zodiacal light removal. 
We do not include the calibration uncertainties (that are on the order of 12\%, \citep{2007PASP..119.1038S}) in our quoted errors. The errors are dominated by those of the zodiacal light model. \\

We can compare the value of the CIB level determined with the TPMs to the mean of the residual map we obtained in Sect. \ref{par:dust+hi}, as the only astrophysical component that should be present in this map is the CIB. We have first to correct the map for the offset. We use 2.05 MJy/sr and find $B_{residual}=0.76$ MJy/sr. This value is in very good agreement with that obtained using the TPMs. They are very close to the predicted CIB level of $B_{160}= 0.63\pm0.02$~MJy/sr by \cite{2011A&A...529A...4B} and to the last Spitzer determination using very deep number counts: \citet{2010A&A...512A..78B} found $B_{160}=0.78^{+0.39}_{-0.15}$ MJy/sr. Moreover, our value is also in very good agreement with that of \citet{2011AA...532A..49B}, who carried out a P(D) analysis in addition to a stacking one (of sources detected at 24~$\mu$m) in Herschel/PACS data at 160 $\mu$m to derive the differential number counts. Extrapolating the counts down to very faint fluxes using a power law, they obtained $B_{160}= 0.72^{+0.19}_{-0.05}$~MJy/sr. We can \textbf{also} compare our determinations with \citet{2009AA...500..763J}, who derived the CIB using ISOPHOT data. They computed linear fits between FIR and \hi~data,  by disregarding several velocity components in \hi~data and for much smaller fields than ELAIS N1 (roughly 25 times smaller). They found  $B_{160}=1.08 \pm 0.32\pm$0.30 MJy/sr in the range 150--180~$\mu$m, where the first error is statistical and the second systematic. Although compatible within the error bars, our determination points to a lower value of the CIB. Our CIB determination benefits from having Spitzer and GBT data that cover a large field, allowing for a more robust measurement of the cirrus contamination and thus of the CIB.\\
We can now combine our CIB mean level measurement at 160~$\mu$m with our CIB anisotropy measurements to compute the CIB at 100~$\mu$m.
\begin{table*}\centering
 \begin{tabular}{*{8}{c}}
 \hline\hline
 Name & Coordinates 	   & TPM value       &$ N_{\hi,local}\times\alpha_{local}$ & $N_{\hi,HVC}\times\alpha_{HVC}$&CIB level                & map           & offset     \\
 \hline
1    & RA=245.20 Dec=54.98 & 2.59$\pm$0.12   &1.40$\pm$0.03	                 & 0.43$\pm$0.01		&0.75$\pm$0.04$\pm$0.12  & 4.78$\pm$0.12 & 2.20$\pm$0.24  \\
2    & RA=243.28 Dec=55.13 & 2.23$\pm$0.12   &1.43$\pm$0.03		         & 0.01$\pm$0.01		&0.78$\pm$0.04$\pm$0.12  & 4.14$\pm$0.12 & 1.91$\pm$0.24 \\
average&                   &                 &                                   &                              &0.77$\pm$0.04$\pm$0.12  &               & 2.05$\pm$0.24\\
 \hline

\end{tabular}
\caption{TPM observations at 160 $\mu$m. The second column gives the coordinates of the TPMs, the third gives the brightness of the TPMs \textbf{(zodiacal-light subtracted)}. Infrared emission of the local and of the HVC are in columns 4 and 5. The sixth column gives CIB levels at 160 $\mu$m. The first error is statistical and the second is systematic. \textbf{It is due to the error on the estimation of the zodiacal light}. The offsets between the scan map and the TPMs are given in the last column. All values are in MJy/sr.}
\label{tab:tpm}
\end{table*}

\subsection{CIB mean estimate at 100 $\mu$m}\label{par:CIB100}
We combine the cirrus-free power spectra measurements at both 100~$\mu$m and 160~$\mu$m with the CIB mean at 160~$\mu$m to derive
the CIB mean level at 100~$\mu$m, following
\begin{equation}
\label{eq_cib_100}
\frac{\sigma^{CIB}_{100 \mu m}}{B^{CIB}_{100 \mu m}}=\frac{\sigma^{CIB}_{160 \mu m}}{B^{CIB}_{160 \mu m}},
\end{equation}
where $\sigma$ is the rms fluctuation in the CIB that can be computed using the measured CIB power spectrum.
Knowing $\sigma^{CIB}_{100 \mu m}$, $\sigma^{CIB}_{160 \mu m}$, and $B^{CIB}_{160 \mu m}$, we can compute $B^{CIB}_{100 \mu m}$ following Eq.~\ref{eq_cib_100}.
We checked using the model of \citet{2012A&A...537A.137P} that Eq. \ref{eq_cib_100} is valid.

The two CIB power spectra determined previously at 100~$\mu$m and 160~$\mu$m were measured with a different flux cut for the bright sources removal. Thus, we first recompute the power spectra for the two maps without masking any bright point sources, in order to have the same point source contamination (i.e. consistent Poisson noise levels). We then perform the same analysis as previously (removal of instrumental noise and cirrus, and division by the power spectrum of the PSF). We compute the 100/160 color of the clustering power spectra using the largest common scales available (0.006$<k<$0.2~arcmin$^{-1}$), obtaining $\sigma^{CIB}_{100 \mu m}/\sigma^{CIB}_{160 \mu m} = 0.31\pm0.1$.
Multiplying the CIB at 160~$\mu$m derived in Sect.~\ref{par:tpm} by $\sigma^{CIB}_{100 \mu m}/\sigma^{CIB}_{160 \mu m}$ leads to a CIB at 100~$\mu$m of $B_{100}^{CIB}=0.24\pm$0.08$\pm$0.04~MJy/sr, where the first error is statistical and the second one is systematic. We note that this value is lower than the mean value of the residual map obtained after the removal of the cirrus component (see Table \ref{tab:only_n1}),  owing to residual zodiacal emission in the IRIS map. An empirical correction of the \citet{2000A&A...354..247L} DIRBE measurements, provided in \citet{2006A&A...451..417D} gives 0.48$\pm$0.21 MJy/sr. Our determination points towards a lower CIB value. By stacking Herschel/PACS maps at 100~$\mu$m at the positions of all 24~$\mu$m galaxies (S(24)$\ge$20$\mu$Jy),  \citet{2010AA...518L..30B} measured a CIB surface brightness of 0.25$\pm$0.02 MJy/sr. Subsequently, \citet{2011AA...532A..49B} improved on these measurements by carrying out a P(D) analysis that allowed them to reach lower fluxes. Extrapolating their differential number counts using a power law, they obtained $B_{100}^{CIB}=0.42^{+0.28}_{-0.06}$~MJy/sr, which is barely compatible with our determination.

\section{Conclusion}
We have presented a new method for removing Galactic cirrus contamination from the power spectrum of CIB anisotropies by using an independent tracer of this cirrus, the \hi~21 cm data. We have computed the far-IR emission of each velocity component of the cirrus and removed it spatially from the maps. The residual map is thus a map of the CIB anisotropies. We have applied this method to MIPS data at 160~$\mu$m in the ELAIS N1 field and recovered the results of \citet{2007ApJ...665L..89L} on intermediate spatial scales where starburst galaxy clustering intervenes. They had derived a linear bias $b \sim 2.4$ with MIPS data at 160~$\mu$m, which probes mainly galaxies around $z\sim1$. We similarly applied this method at 100~$\mu$m and detected for the first time the correlated anisotropies at this wavelength. An analysis of these CIB anisotropy power spectra was not an objective of this paper, and we refer the reader to \citet{2012A&A...537A.137P}, who presented a clustering model of star-forming galaxies. \\
We have shown that the cirrus removal using \hi~data is the most efficient available method for removing the cirrus contamination accurately. It has also been successfully applied to Planck data \citep{2011A&A...536A..18P} and will be applied to Herschel data. We caution the reader that it is incorrect to use the 100 $\mu$m map of \citet{1998ApJ...500..525S} as a cirrus tracer as this map  contains the anisotropies of the CIB.\\
We have used absolute measurements of the brightness on small regions of the sky (total power modes) with different cirrus contributions to derive the CIB level. Making use of our emissivities, we were able to compute the cirrus brightness in two of these regions, and derive the CIB level at 160~$\mu$m. We found $B_{160}=0.77\pm$0.04$\pm$0.12 MJy/sr (the first error is statistical and the second systematic). In addition, using our measured CIB correlated anisotropies at 100~$\mu$m we computed the CIB anisotropy color, $B_{100}/B_{160}= 0.31\pm0.1$. This color measurement is free of cirrus and zodiacal light uncertainties, since the former had been removed from both power spectra and the latter is a constant that has no influence on power spectra. Using this color and the CIB measured at 160~$\mu$m, we derived the CIB at 100~$\mu$m $B_{100}=0.24\pm 0.08\pm0.04$ MJy/sr where the first error is statistical and the second systematic. These CIB measurements are the most accurate measurements based on a diffuse emission analysis. Comparing those measurements with an extrapolation of galaxy number counts, we have found no evidence of an unknown contribution to the CIB, in contrast to \citet{2011ApJ...737....2M}.

\bibliographystyle{aa}

\bibliography{Biblio}

\end{document}